\documentclass[11pt,draftcls,onecolumn]{IEEEtran}

\ifCLASSINFOpdf
   \usepackage[pdftex]{graphicx}
\else
   \usepackage[dvips]{graphicx}
   \DeclareGraphicsExtensions{.eps}
\fi

\usepackage[cmex10]{amsmath}
\usepackage{url}
\usepackage{amssymb}
\usepackage{pifont}
\usepackage{color}

\newtheorem{remark}{Remark}
\newcommand{\bi}[1]{\textbf{\textit{#1}}}

\newcommand{\bb}[1]{\textbf{#1}}

\newcommand{\NN}{{\mathbb N}}

\newcommand{\CC}{{\mathbb C}}
\newcommand{\ZZ}{{\mathbb Z}}
\newcommand{\RR}{{\mathbb R}}

\newcommand{\Ree}[0]{\mathrm{Re}}

\def\adots{\mathinner{\mkern 2mu\raise 1pt\hbox{.}\mkern 3mu\raise 4pt\hbox{.}\mkern 1mu\raise 7pt\hbox{.}}}

\begin{document}

\title{Optimization of Synthesis Oversampled Complex Filter Banks}

\author{J\'er\^ome~Gauthier,~\IEEEmembership{Student Member,~IEEE},
        Laurent~Duval,~\IEEEmembership{Member,~IEEE}
        and~\\Jean-Christophe~Pesquet,~\IEEEmembership{Senior Member,~IEEE}
\thanks{Copyright (c) 2008 IEEE. Personal use of this material is permitted. However, permission to use this material for any other purposes must be obtained from the IEEE by sending a request to pubs-permissions@ieee.org.}
\thanks{L. Duval is with the Institut Fran\c{c}ais du P\'etrole, IFP, Technology, Computer
Science and Applied Mathematics Division, 1 et 4, avenue de Bois-Pr\'eau 
92852 Rueil-Malmaison, France. E-mail: laurent.duval@ifp.fr.}
\thanks{J. Gauthier and J.-C. Pesquet are with the Institut Gaspard Monge and CNRS-UMR
8049, Universit\'e de Paris-Est, 77454 Marne-la-Vall\'ee
Cedex 2, France. E-mail: \{jerome.gauthier,jean-christophe.pesquet\}@univ-paris-est.fr.}
}


\maketitle

\begin{abstract}
An important issue with oversampled FIR analysis filter banks (FBs) is 
to determine inverse synthesis FBs, when they exist. 
Given any  complex oversampled FIR analysis FB, we first provide an algorithm to determine whether there exists an inverse FIR synthesis system. We also provide a method to ensure the  Hermitian symmetry property on the synthesis
side, which is serviceable to processing real-valued signals.
As an invertible analysis scheme corresponds to a redundant 
decomposition, there is no unique inverse FB. 
Given a particular solution, we parameterize the whole family of inverses through a null space projection. The resulting reduced parameter set simplifies design procedures, since the perfect reconstruction constrained optimization problem is recast as an unconstrained optimization problem.
The design of optimized synthesis FBs based on time or frequency localization criteria
is then investigated, using a simple yet efficient gradient algorithm.
\end{abstract}



\section{Introduction}
Since the 70s, filter banks (FBs) have become a central tool in signal/image processing and communications: lapped  or discrete wavelet transforms can be viewed  as instances of FB structures. Likewise, oversampled FBs (OFBs) constitute an extensively studied instance with remaining open questions. Their development came along under a variety of different appellations, to name a few: general Analysis-Synthesis Systems \cite{Kellermann_W_1988_icassp_ana_dmscae}, DFT (discrete Fourier transform) with stack-shift capability, 
Overlap-Add or Generalized  DFT, underdecimated systems, oversampled harmonic modulated filter banks \cite{Bolcskei_H_1998_tcas2_ove_cmfbpr,Labeau_F_2005_tsp_ove_fbecctinc},  complex lapped transforms \cite{Young_R_1993_tip_fre_dmeclt}, generalized lapped pseudo-biorthogonal transform, \emph{etc}. 

In a more generic form, OFBs have received a considerable attention both theoretically and in many applications, in the past ten years, following their association with specific types of frames \cite{Bolcskei_H_1998_tcas2_ove_cmfbpr,Bolcskei_H_1998_tsp_fra_taofb,Cvetkovic_Z_1998_tsp_ove_fb}.
Their design flexibility, improved frequency selectivity and increased robustness to noise and aliasing distortions have made them useful for subband adaptive filtering in audio processing \cite{Malvar_H_1999_icassp_mod_cltaap}, noise shaping \cite{Bolcskei_H_1997_icassp_ove_fbonsdfna}, denoising \cite{Labeau_F_2005_tsp_ove_fbecctinc},  multiple description coding  \cite{Goyal_V_2001_acha_qua_fee}, echo cancellation \cite{Dumitrescu_B_2006_jasp_sim_dldonprgdft}, multiple antenna code design \cite{Hassibi_B_2001_tit_rep_thrmacd}, channel equalization 
\cite{Pun_K_2007_icassp_des_ofbbwloqpske,Ihalainen_T_2007_jasp_cha_efbbmmwc,Johansson_H_2007_jasp_fle_fbrnvocmfb}
 or  channel coding \cite{Weiss_S_2006_jasp_par_ofbdcc}. 

Two major problems arise when resorting to OFBs: (\emph{i}) the existence of an inverse for the analysis OFB achieving perfect reconstruction (PR) 
and (\emph{ii}) the determination of  an ``optimal'' synthesis FB. Since the additional degrees of freedom gained through redundancy 
may increase the design complexity, 
several works have focused on FBs modulated with a single \cite{Cvetkovic_Z_1998_tsp_tig_whfl2r,Yui_K_2004_spl_mul_doudftfb} or  multiple windows \cite{Ueng_N_1996_icassp_fra_obvwft}. More general formulations are based on factorizations of OFB polyphase   representations with additional constraints (restricted oversampling ratios, symmetry, realness or filter length) into a lattice  \cite{VonBorries_R_2001_icassp_fil_ro,VonBorries_R_2004_icassp_lin_pofb,Gan_L_2003_tsp_ove_lpprftlsp,Tanaka_T_2006_tsp_dir_doprfirfbfof} or a lifting structure \cite{Riel_B_2004_midwest_lif_bdiopmfb}.
 Constructions  with near perfect reconstruction (relaxing the PR property) have also been proposed \cite{Harteneck_M_1999_tcas2_des_nprofbsaf,Wilbur_M_2003_tr_eff_onprgdftfb,Dumitrescu_B_2006_jasp_sim_dldonprgdft,Hermann_D_2007_icassp_win_bpfdhofbaa}. In \cite{Kalker_T_1995_icassp_gro_btmms,Park_H_1997_mssp_gro_bmfirms,Zhou_J_2005_spie-wav_mul_ofb},  more  involved algebraic tools (such as Gr{\"o}bner bases) have also been employed. Recently, Chai \emph{et al.} have proposed a design based on FB state-space representations  \cite{Chai_L_2007_tsp_fra_tbadofbdcm}. The design may use different kinds of optimization
criteria based on filter regularity or more traditional cost functions based on filter shape (subband attenuation \cite{Gan_L_2003_tsp_ove_lpprftlsp,Dumitrescu_B_2006_jasp_sim_dldonprgdft}, coding gain \cite{Labeau_F_2005_spl_syn_fdcgofb}). Most of those synthesis FB designs rely on minimum-norm solutions. An interesting approach combining the latters with a null space method was successfully pursued  by Mansour \cite{Mansour_M_2007_spl_opt_odftfb} for synthesis window shape optimization in a modulated DFT FB.

Within the compass  of the proposed work is a relatively generic construction and optimization of  oversampled synthesis filter banks with Finite Impulse Response (FIR) properties at both the analysis and synthesis sides. We can additionally impose a practically useful Hermitian symmetry on the synthesis side. This work extends the results given in two previous conference papers  \cite{Gauthier_J_2006_picassp_low_rolta3Dsdf,Gauthier_J_2007_picassp_ove_iclto}. A special case has been judiciously devised in \cite{Tanaka_T_2006_tsp_dir_doprfirfbfof}, for specific filter length and redundancy factor allowing closed form expressions for two design criteria.
In Section \ref{sec:formulpolyph1d} we recall the polyphase notation used throughout this paper. Given arbitrary FIR complex oversampled analysis FB, we first describe in Section \ref{sec:exist} a simple algorithm to test whether it is FIR invertible or not, based on known results on polynomial matrices \cite{Fornasini_E_1997_mssp_pol_mamsa,Park_H_2003_jsc_sym_csp}.  
The standard  Moore-Penrose pseudo-inverse (PI) solution \cite{BenIsrael_A_2003_book_gen_ita} is studied in Section \ref{subsec:comp}.
In Section  \ref{sec:invsymcase}, a method is supplied to enforce an Hermitian symmetric FB, which is useful for real data analysis, processing and synthesis. In Section \ref{sec:tspoptim}, the problem of the optimal design of the synthesis
FB is addressed. Although optimization can be
studied both on the analysis and synthesis sides \cite{Tanaka_T_2004_tsp_gen_lpbtolpprfbls,Komeiji_S_2007_ncsp_dir_doprbfirfbrc}, we consider here a given analysis FB and work on the synthesis side. We derive in Section \ref{sec:dimred} an efficient parameter set size reduction for this purpose. 
Using time or frequency localization criteria, we then reformulate in Section \ref{sec:optsol} the constrained optimization problem as an unconstrained one for both the  general and Hermitian symmetric cases. After describing  the optimization process, we illustrate, in Section~\ref{sec:tspexamples}, the different methods proposed for the inversion and optimization on three classical oversampled real and complex FB types.

\section{Problem statement}
\label{sec:formulpolyph1d}

\subsection{Notations}

\begin{center}
\begin{figure}[ht]
\begin{center}
 \setlength{\unitlength}{3066sp}%
\begingroup\makeatletter\ifx\SetFigFont\undefined%
\gdef\SetFigFont#1#2#3#4#5{%
  \reset@font\fontsize{#1}{#2pt}%
  \fontfamily{#3}\fontseries{#4}\fontshape{#5}%
  \selectfont}%
\fi\endgroup%
\begin{picture}(4994,2004)(1014,-7273)
\put(3746,-5866){\rotatebox{270.0}{\makebox(0,0)[lb]{\smash{{\SetFigFont{9}{10.8}{\rmdefault}{\mddefault}{\updefault}{ Processing}%
}}}}}
\thicklines
{ \put(1296,-6406){\line(-1, 0){360}}
}%
\thinlines
{ \multiput(3566,-7261)(9.00000,0.00000){56}{\makebox(2.1453,15.0169){\SetFigFont{5}{6}{\rmdefault}{\mddefault}{\updefault}.}}
\multiput(3566,-5281)(9.00000,0.00000){56}{\makebox(2.1453,15.0169){\SetFigFont{5}{6}{\rmdefault}{\mddefault}{\updefault}.}}
\multiput(3566,-7261)(0.00000,9.00000){221}{\makebox(2.1453,15.0169){\SetFigFont{5}{6}{\rmdefault}{\mddefault}{\updefault}.}}
\multiput(4061,-7261)(0.00000,9.00000){221}{\makebox(2.1453,15.0169){\SetFigFont{5}{6}{\rmdefault}{\mddefault}{\updefault}.}}
}%
\thicklines
{ \put(1296,-5641){\vector( 1, 0){180}}
\put(1296,-5641){\line( 0,-1){810}}
}%
{ \put(1296,-6181){\vector( 1, 0){180}}
}%
{ \put(5591,-5641){\line( 1, 0){225}}
\put(5816,-5641){\line( 0,-1){810}}
}%
{ \put(5591,-6181){\line( 1, 0){225}}
}%
{ \put(5816,-6406){\vector( 1, 0){270}}
}%
{ \put(5636,-6901){\line( 1, 0){180}}
\put(5816,-6901){\line( 0, 1){180}}
}%
{ \put(1296,-6901){\vector( 1, 0){180}}
\put(1296,-6901){\line( 0, 1){180}}
}%
\put(1521,-6361){\framebox(630,360){}}
\put(1521,-5821){\framebox(630,360){}}
\put(4106,-5641){\vector( 1, 0){225}}
\put(4106,-6181){\vector( 1, 0){225}}
\put(4106,-6901){\vector( 1, 0){225}}
\put(4331,-7081){\framebox(450,360){}}
\put(4331,-6361){\framebox(450,360){}}
\put(4331,-5821){\framebox(450,360){}}
\put(4781,-5641){\vector( 1, 0){180}}
\put(4781,-6181){\vector( 1, 0){180}}
\put(4781,-6901){\vector( 1, 0){180}}
\put(4961,-7081){\framebox(630,360){}}
\put(4961,-6361){\framebox(630,360){}}
\put(4966,-5821){\framebox(625,360){}}
\put(3196,-6766){\makebox(0,0)[b]{\smash{{\SetFigFont{9}{10.8}{\familydefault}{\mddefault}{\updefault}$y_{M-1}(n)$}}}}
\put(3196,-6046){\makebox(0,0)[b]{\smash{{\SetFigFont{9}{10.8}{\familydefault}{\mddefault}{\updefault}$y_1(n)$}}}}
\put(3196,-5506){\makebox(0,0)[b]{\smash{{\SetFigFont{9}{10.8}{\familydefault}{\mddefault}{\updefault}$y_0(n)$}}}}
\put(2556,-5686){\makebox(0,0)[b]{\smash{{\SetFigFont{9}{10.8}{\familydefault}{\mddefault}{\updefault}$\downarrow\!N$}}}}
\put(2556,-6226){\makebox(0,0)[b]{\smash{{\SetFigFont{9}{10.8}{\familydefault}{\mddefault}{\updefault}$\downarrow\!N$}}}}
\put(2556,-6946){\makebox(0,0)[b]{\smash{{\SetFigFont{9}{10.8}{\familydefault}{\mddefault}{\updefault}$\downarrow\!N$}}}}
\put(1836,-5686){\makebox(0,0)[b]{\smash{{\SetFigFont{9}{10.8}{\familydefault}{\mddefault}{\updefault}$H_0$}}}}
\put(1836,-6226){\makebox(0,0)[b]{\smash{{\SetFigFont{9}{10.8}{\familydefault}{\mddefault}{\updefault}$H_1$}}}}
\put(1836,-6946){\makebox(0,0)[b]{\smash{{\SetFigFont{9}{10.8}{\familydefault}{\mddefault}{\updefault}$H_{M-1}$}}}}
\put(1296,-6676){\makebox(0,0)[b]{\smash{{\SetFigFont{9}{10.8}{\familydefault}{\mddefault}{\updefault}$\vdots$}}}}
\put(1026,-6271){\makebox(0,0)[b]{\smash{{\SetFigFont{9}{10.8}{\familydefault}{\mddefault}{\updefault}$\textbf{x}_n$}}}}
\put(4556,-5686){\makebox(0,0)[b]{\smash{{\SetFigFont{9}{10.8}{\familydefault}{\mddefault}{\updefault}$\uparrow\!N$}}}}
\put(2781,-6901){\vector( 1, 0){740}}
\put(4556,-6946){\makebox(0,0)[b]{\smash{{\SetFigFont{9}{10.8}{\familydefault}{\mddefault}{\updefault}$\uparrow\!N$}}}}
\put(5276,-5686){\makebox(0,0)[b]{\smash{{\SetFigFont{9}{10.8}{\familydefault}{\mddefault}{\updefault}$\widetilde{H}_0$}}}}
\put(5276,-6226){\makebox(0,0)[b]{\smash{{\SetFigFont{9}{10.8}{\familydefault}{\mddefault}{\updefault}$\widetilde{H}_1$}}}}
\put(5276,-6946){\makebox(0,0)[b]{\smash{{\SetFigFont{9}{10.8}{\familydefault}{\mddefault}{\updefault}$\widetilde{H}_{M-1}$}}}}
\put(5816,-6676){\makebox(0,0)[b]{\smash{{\SetFigFont{9}{10.8}{\familydefault}{\mddefault}{\updefault}$\vdots$}}}}
\put(4556,-6226){\makebox(0,0)[b]{\smash{{\SetFigFont{9}{10.8}{\familydefault}{\mddefault}{\updefault}$\uparrow\!N$}}}}
\put(5816,-6454){\makebox(0,0)[b]{\smash{{\SetFigFont{9}{10.8}{\familydefault}{\mddefault}{\updefault}$\oplus$}}}}
\put(2781,-5641){\vector( 1, 0){740}}
\put(2781,-6181){\vector( 1, 0){740}}
\put(2331,-5821){\framebox(450,360){}}
\put(2331,-6361){\framebox(450,360){}}
\put(2331,-7081){\framebox(450,360){}}
\put(2151,-6181){\vector( 1, 0){180}}
\put(2151,-6901){\vector( 1, 0){180}}
\put(2151,-5641){\vector( 1, 0){180}}
\put(1521,-7081){\framebox(630,360){}}
\end{picture}%
 \caption{Oversampled $M$-channel filter bank.}
 \label{fig:fbdiagram}
\end{center}
\end{figure}
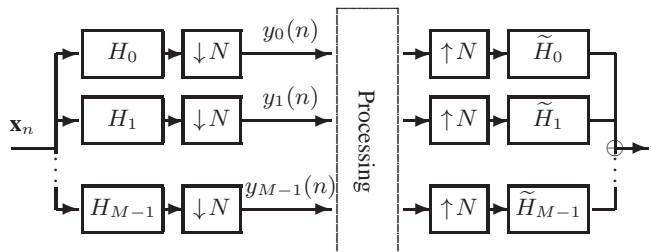 
\end{center}

Lapped transforms \cite{Malvar_H_1992_book_sig_plt} were introduced in \cite{Cassereau_P_1989_tcom_enc_iblot} to avoid blocking artifacts in audio processing. Similarly for images, they reduce tiling effects produced by classical block transforms (as can be seen in the JPEG image compression format). Lapped transforms belong to the class of FBs, such as the one represented in Figure~\ref{fig:fbdiagram}, with a decimation factor $N$ smaller than the length of each filter. The filters, whose impulse responses are denoted by $(h_{i})_{0 \leq i < M}$, are supposed of finite length $kN$ with $k$ an integer greater than or equal to 2. We therefore consider $k$ overlapping blocks of size $N$.

A signal $(x(n))_{n\in \mathbb{Z}}$ is decomposed by $M$ filters; since the decimation factor is $N$, the overall redundancy of the transform is $M/N=k'$. In this paper, we investigate the oversampled case, \emph{i.e.} $k'>1$.
The $M$ outputs of the analysis FB are denoted by $(y_{i}(n))_{0\leq i <M}$. 
With these notations, the outputs of the analysis FB are expressed, for all $i\in \{0,\ldots,M-1\}$
and $n\in \mathbb{Z}$, as
\begin{equation}
\label{eqdef}
y_{i}(n)  = \sum_{p} h_i(p) x(Nn-p)  = \sum_{\ell}\sum_{j=0}^{N-1} h_i(N\ell+j) x(N(n-\ell)-j).
\end{equation} 

\subsection{Polyphase formulation}

Let $\bi{H}(\ell)=\left[h_{i}(N\ell+j)\right]_{0\leq i < M,0 \leq j <N}$, $\ell\in\left\{0,\ldots,k-1\right\}$
be the $k$ polyphase matrices obtained from the impulse responses of the analysis filters. We also define the polyphase vector from the input signal $x(n)$:
$\forall n\in \mathbb{Z}, \; \bb{x}(n)=\left(x(Nn-j)\right)_{0\leq j <N}$, leading to concisely rewriting \eqref{eqdef} into a convolutive form:
\begin{equation}
\label{eqdef2}
\bb{y}(n)= (y_0(n),\ldots,y_{M-1}(n))^\top= \sum^{k-1}_{\ell=0}\bi{H}(\ell)\bb{x}(n-\ell)= 
\left(\bi{H}*\bb{x}\right)(n),
\end{equation}
where $^\top$ is the transpose operator. Thus, \eqref{eqdef2} can be reexpressed as:
$\bb{y}[z]=
\bi{H}[z]\bb{x}[z]$, 
where  
$\bi{H}[z]=\sum^{k-1}_{\ell=0}\bi{H}(\ell)z^{-\ell}$ 
is the $M\times N$ polyphase transfer matrix of the analysis FB and $\bb{x}[z]$ (resp. $\bb{y}[z]$) is the $z$-transform of $(\bb{x}(n))_{n\in \mathbb{Z}}$ (resp. $(\bb{y}(n))_{n\in \mathbb{Z}}$).

\subsection{Synthesis FB}
The polyphase transfer matrix of the synthesis FB:
$\widetilde{\bi{H}}[z]=\sum_{\ell}\widetilde{\bi{H}}(\ell)z^{-\ell}$, satisfies:
\begin{equation}
\widetilde{\bb{x}}[z]=\widetilde{\bi{H}}[z]\bb{y}[z],
\label{eq:reconpolygen}
\end{equation}
where the polyphase vector of the ouput signal of the synthesis FB $(\widetilde{\bb{x}}(n))_{n\in\mathbb{Z}}$ is defined similarly to  $(\bb{x}(n))_{n\in\mathbb{Z}}$. We deduce from \eqref{eq:reconpolygen} that:
\begin{equation}
\label{eqid1}
\forall n\in\ZZ,\forall i\in\left\{0,...,N-1\right\},\quad \widetilde{x}(nN-i)=\sum^{M-1}_{j=0}\sum^{\infty}_{\ell=-\infty}\widetilde{H}_{i,j}(n-\ell)y_{j}(\ell), 
\end{equation}
where $\widetilde{\bi{H}}(\ell)=\left(\widetilde{H}_{i,j}(\ell)\right)_{0\le i<N,0\le j<M}$. Expressing \eqref{eqid1} with impulse responses,
we  can write: for every $i\in\left\{0,...,N-1 \right\}$ and $n\in\ZZ$,
\begin{equation}
\label{eqid2}
\widetilde{x}(nN-i)=\sum^{M-1}_{j=0}\sum^{\infty}_{\ell=-\infty}\widetilde{h}_{j}(N(n-\ell)-i)y_{j}(\ell)
\end{equation}
which, by identifying \eqref{eqid1} and \eqref{eqid2}, allows us to deduce that
\begin{equation}
\label{eqhtildeimpul}
\widetilde{\bi{H}}(\ell)=\left[\widetilde{h}_{j}(N\ell-i)\right]_{0\leq i < N,0 \leq j <M},\quad \ell\in\mathbb{Z}.
\end{equation}
These expressions hold for any oversampled FIR FB.

\section{Inversion}

\subsection{Invertibility of an analysis FB}
\label{sec:exist}

This work being focused on the construction of FIR synthesis filters, a preliminary point is the verification of the given analysis FB FIR invertibility. The polyphase representation of FBs offers the advantage of relating  the perfect reconstruction property to the invertibility of the polyphase transfer matrix \cite{Vaidyanathan_P_1993_book_mul_sfb}. The latter matrix belongs to the ring $\CC[z,z^{-1}]^{M\times N}$ of Laurent polynomial matrices of dimensions $M \times N$. We emphasize that we do not look for any inverse MIMO filter, but for an inverse
\textit{polynomial} matrix in $\CC[z,z^{-1}]^{N\times M}$ instead. In other words, we aim at obtaining a (non-necessarily causal) FIR synthesis FB.

A first answer to this FIR invertibility problem can be conveyed through the study of the Smith McMillan form of a polynomial matrix 
\cite{Kailath_T_1980_book_lin_s,Vaidyanathan_P_1993_book_mul_sfb}, but unfortunately this decomposition is quite costly. Park, Kalker and Vetterli also devised a method using Gr\"{o}bner bases \cite{Park_H_1997_mssp_gro_bmfirms} to study the invertibility of polynomial matrices which is applicable to the general multidimensional case. We describe here an alternative cost-effective method in the one-dimensional case.
The following result gives a necessary and sufficient condition for a matrix to be left invertible, and thus, for the existence of such an inverse system: let $\bi{H}[z]\in\CC[z,z^{-1}]^{M\times N}$ be a polynomial matrix with $M > N$. The following conditions are equivalent:
\begin{enumerate}
  \item $\bi{H}[z]$ is ``coprime'', which means that the determinants of the maximum minors (sub-matrices of size $N\times N$) are mutually relatively prime.
	\item $\bi{H}[z]$ is left invertible in the sense that there exists $\widetilde{\bi{H}}[z]\in\CC[z,z^{-1}]^{N\times M}$ such that  $\widetilde{\bi{H}}[z]\bi{H}[z]=\bi{I}_{N}$.
\end{enumerate}
A proof of this result can be found in \cite{Fornasini_E_1997_mssp_pol_mamsa} for instance.

The first condition is directly applicable in practice to resolve the left invertibility of the polyphase transfer matrix.
Using the following procedure, we can check numerically whether this condition is satisfied: 
\begin{dingautolist}{172}
	\item Extract  a maximal sub-matrix $\bi{H}_{e}[z]$ of $\bi{H}[z]$.
	\item Compute $\det(\bi{H}_{e}[z])$, and determine its set of roots $\mathcal{S}_{e}$.
  \item Consider another maximal sub-matrix. Remove from $\mathcal{S}_{e}$ the elements which are not roots of the determinant of this sub-matrix.
	\item Repeat step \ding{174} until $\mathcal{S}_{e} = \emptyset$ or all maximal sub-matrices have been extracted.
	\item If $\mathcal{S}_{e} = \emptyset$
then the po\-ly\-pha\-se transfer matrix is left-invertible; otherwise, it is not.
\end{dingautolist}

The corresponding algorithm is easily implemented, leading to 
extract the roots of a single polynomial and check 
the roots of at most $\binom{M}{N}-1=\frac{M!}{N!(M-N)!}-1$ polynomials. If the polyphase matrix is left invertible, the number of considered polynomials in practice is usually much smaller than $\binom{M}{N}-1$, this bound being   reached only when the matrix is not invertible. 
Note that in the case of causal filters (i.e. both $\bi{H}[z]$ and $\widetilde{\bi{H}}[z]$ are polynomial matrices in $\CC[z^{-1}]^{N\times M}$), simpler invertibility conditions exist by invoking the so-called \emph{column-reduced} property \cite{Forney_G_1975_siam-co_min_brvsamls,Inouye_Y_2002_tcas_sys_tfbefmimocs}. 
Also notice that, one of the advantages of this algorithm over other methods is that it can be fully numerically implemented. 

\subsection{Computation of an inverse FB}
\label{subsec:comp}

The method proposed in Section~\ref{sec:exist} only guarantees the existence of a left-inverse, corresponding to an FIR synthesis FB. Since it does not provide a constructive expression, we now perform the actual computation of an inverse polyphase transfer matrix. We assume hereafter that $\bi{H}[z]$ was proven to be FIR left invertible.

Since the goal is to achieve PR, we search for a matrix $\widetilde{\bi{H}}[z]$ in $\CC[z,z^{-1}]^{N\times M}$ such that $\widetilde{\bi{H}}[z]\,\bi{H}[z]=\bi{I}_{N}$ and there exists $(p_1,p_2)\in\NN^2$  such that the polyphase transfer function of the synthesis FB reads: $\widetilde{\bi{H}}[z]=\sum^{p_{2}}_{\ell=-p_{1}}\widetilde{\bi{H}}(\ell)z^{-\ell}$. The resulting overlapping factor of the synthesis filters is $p=p_{1}+p_{2}+1$. 
When working with Laurent polynomial matrices, these integers $p_{1}$ and $p_{2}$ are \emph{a priori} unknown, whereas with polynomial matrices a bound exists \cite{Forney_G_1975_siam-co_min_brvsamls}. By rewriting the PR property in block convolutional form, we get the following linear system:
\begin{equation}
{\cal{H}}\widetilde{{\cal{H}}}={\cal{U}}
\label{eq:HtildeHU}
\end{equation}
where 
\begin{equation}
{\cal{U}}^\top= \left[\bi{0}_{N,p_{1}N},\;\bi{I}_{N},\;\bi{0}_{N,(p_{2}+k-1)N}\right]\in\RR^{N\times (k+p-1)N},\,\,\widetilde{{\cal{H}}}^\top=\left[\widetilde{\bi{H}}(-p_{1}),\cdots,\widetilde{\bi{H}}(p_{2})\right]\in\CC^{N\times pM},\label{eq:defHtildecur}
\end{equation}
\begin{equation}
\mbox{and }\quad
\label{eq:defmathtilde}
{\cal{H}}^\top=\left(\begin{array}{ccccc}
 \bi{H}(0) & \cdots & \bi{H}(k-1) &  &  0\\
  &  \ddots & &\ddots &  \\
  0 &   & \bi{H}(0) &  \cdots & \bi{H}(k-1) \\
  \end{array}\
\right)\in\CC^{pM\times(k+p-1)N}.
\end{equation}

As already mentioned, $p_{1}$ and $p_{2}$ are unknown, but since the system \eqref{eq:HtildeHU} is supposed invertible,  at least a couple of integers $(p_{1},p_{2})$ solving the system exists. The values of $p_1$ and $p_2$ are actually obtained by increasing the value of $p$  and looking for every couple satisfying $p=p_{1}+p_{2}+1$, starting with  $p=1$. 
Hence, for a given $p$, we consider all $(p_{1},p_{2})$ in $\left\{(p-1,0),(p-2,1),...,(0,p-1)\right\}$. The first $p$ allowing a Moore-Penrose pseudo-inverse \cite{Penrose_R_1955_pcps_gen_im} solution to \eqref{eq:HtildeHU} provides an inverse polyphase transfer matrix of minimum order.

\subsection{Hermitian symmetric case}
\label{sec:invsymcase}

\subsubsection{Symmetry conditions}

It is well known that the Fourier transform of a real signal is Hermitian Symmetric (HS): its frequency decomposition is symmetric for the real part and anti-symmetric for the imaginary part. Conversely, if the coefficients are HS in the frequency domain, then the reconstructed signal is real. 
This property is very useful for real data filtering, which often consists of removing or thresholding coefficients in the frequency domain before reconstructing. Securing the reconstruction
of real-valued signals from the transformed coefficients
is thus a desirable property. 
In this section, we study the HS case and its effects on the methods proposed in the previous sections. 
 
The HS property in the synthesis filters is satisfied provided that,
 considering any symmetric subband indices $j_{f}\in\left\{0,...,M-1\right\}$
and $M-1-j_{f}$, for any coefficients $(y_{i}(n))_{0\leq i<M}$
such that $y_{i}(n)=0$ 
if $(i,n)\neq(j_{f},n_{f})$ or $(i,n)\neq(M-1-j_{f},n_{f})$ with $n_{f}\in\ZZ$ 
and, such that $y_{j_{f}}(n_{f})=\overline{y_{M-1-j_{f}}(n_{f})}$, 
a real-valued signal is reconstructed. 
The reconstructed signal reads:
\begin{align*}
\widetilde{x}(m)&=\sum^{M-1}_{j=0}\sum^{\infty}_{\ell=-\infty}\widetilde{h}_{j}(m-N\ell)y_{j}(\ell)=\widetilde{h}_{j_{f}}(m-n_{f}N)y_{j_{f}}(n_{f})+\widetilde{h}_{M-1-j_{f}}(m-n_{f}N)\overline{y_{j_{f}}(n_{f})}.
\end{align*}
A necessary and sufficient condition for $\widetilde{x}(m)\in\RR$ for all $y_{j_{f}}(n_{f})\in\CC$, is that  $\widetilde{h}_{j_{f}}(m-n_{f}N)=\overline{\widetilde{h}_{M-1-j_{f}}(m-n_{f}N)}.$
This condition must be verified for any couple of integers $(j_{f},n_{f})$. 
The condition on the synthesis filter is then: 
 $\forall j\in\left\{0,...,M-1\right\}\mbox{ and }\forall n\in\ZZ,\quad\widetilde{h}_{j}(n)=\overline{\widetilde{h}_{M-1-j}(n)}.$ 
Using \eqref{eqhtildeimpul}, we rewrite the condition as 
\begin{equation}\forall \ell\in\left\{-p_{1},...,p_{2} \right\},\quad\widetilde{\bi{H}}(\ell)= \overline{\widetilde{\bi{H}}(\ell)}\bi{J}_{M}.\label{eq:cond1}
\end{equation}
where $\bi{J}_{M}$ is the $M\times M$ counter-identity matrix:
$$\bi{J}_{M}=\left(\begin{array}{ccc}
 0 &  &  1\\
  &  \adots &  \\
  1 &   & 0 \\
  \end{array}\
\right).$$
We have supposed here that the transformed coefficients of a real signal exhibit the HS property. In other words, for any real signal
 $(x(n))_{n\in\ZZ}$, and for any couple $(j_{f},n_{f})\in\left\{0,...,M-1\right\}\times\ZZ$, the output of the analysis FB verifies:
$y_{j_{f}}(n_{f})=\overline{y_{M-j_{f}-1}(n_{f})}. $ 
This condition can be rewritten:
$$\sum_{m}h_{j_{f}}(m)x(Nn_{f}-m)=\sum_{m}\overline{h_{M-j_{f}-1}(m)}x(Nn_{f}-m). $$
Considering a zero input signal $x$ except for one sample, we deduce that
$h_{j_{f}}(n)=\overline{h_{M-j_{f}-1}}(n),$ 
which is equivalent to
\begin{equation}\bi{H}(\ell)=\bi{J}_{M} \overline{\bi{H}(\ell)}. \label{eq:cond2}
\end{equation}
Hence, if the analysis FB verifies Condition \eqref{eq:cond2}, then the coefficients after decomposition satisfy the HS property. 

\begin{remark}Consider an invertible HS analysis FB. By inserting \eqref{eq:cond2} in the PR condition we get:
$$ \delta_{\ell}\bi{I}_{N}=\sum^{\min(p_{2},\ell)}_{s=\max(\ell-k+1,-p_{1})}\overline{\widetilde{\bi{H}}(s)}\bi{J}_{M} \bi{H}(\ell-s).$$
It implies that if $\widetilde{{\cal{H}}}=\left[\widetilde{\bi{H}}(-p_{1}),\cdots,\widetilde{\bi{H}}(p_{2})\right]^\top$ is a solution of the linear system \eqref{eq:HtildeHU} then, under the HS hypothesis on the analysis FB, 
$\widetilde{{\cal{H}}_{2}}=\left[\overline{\widetilde{\bi{H}}(-p_{1})}\bi{J}_{M},\cdots,\overline{\widetilde{\bi{H}}(p_{2})}\bi{J}_{M}\right]^\top$
is also a solution of the linear system. Finally, it follows that the sum:
$\widetilde{{\cal{H}}_{0}}=\frac{1}{2}(\widetilde{{\cal{H}}}+\widetilde{{\cal{H}}_{2}})$ is also a solution. Moreover, this solution verifies Condition~\eqref{eq:cond1} by construction.\\
In other words, we have proved that an invertible HS analysis FB admits at least one HS synthesis FB.
\end{remark}

\subsubsection{Construction method}
\label{sec:invsymcasemeth2}

We suppose here that the analysis FB was proven invertible and that the matrices $\bi{H}(\ell)$ satisfy Condition \eqref{eq:cond2}. Our objective is to build a synthesis FB possessing both the PR and HS properties.

\paragraph{First case: $M$ is even}
\label{sssec:Mpair}
First, we rewrite Conditions \eqref{eq:cond1} and \eqref{eq:cond2}:
\begin{align*}
(\ref{eq:cond1})&\Leftrightarrow\forall \ell\in\left\{-p_{1},...,p_{2}\right\}, \left\{ \begin{array}{l} \widetilde{\bi{H}}(\ell)=\left[\widetilde{\bi{H}_{1}}(\ell), \widetilde{\bi{H}_{2}}(\ell)\right],\, \widetilde{\bi{H}_{1}}\in \CC^{N\times M'}\mbox{ and }\widetilde{\bi{H}_{2}} \in \CC^{N\times M'}, \\
\widetilde{\bi{H}_{1}}(\ell)= \overline{\widetilde{\bi{H}_{2}}(\ell)}\bi{J}_{M'}\\
\end{array}
\right.\\
(\ref{eq:cond2})&\Leftrightarrow\forall \ell\in\left\{0,...,k-1\right\}, \left\{ \begin{array}{l} \bi{H}(\ell)=\left(
\begin{array}{c}
\bi{H}_{1}(\ell)\\ 
\bi{H}_{2}(\ell)\\
\end{array}
\right),\, \bi{H}_{1}\in \CC^{M'\times N}\mbox{ and }\bi{H}_{2} \in \CC^{M'\times N}, \\
\bi{H}_{1}(\ell)= \bi{J}_{M'}\overline{\bi{H}_{2}(\ell)}\\
\end{array}
\right.
\end{align*}
with $M'=M/2$. 
Combining these conditions and the PR property, we get:
\begin{align}
\delta_{\ell}\bi{I}_{N}
&=\sum^{\min(p_{2},\ell)}_{s=\max(\ell-k+1,-p_{1})}\widetilde{\bi{H}_{1}}(s)\bi{H}_{1}(\ell-s) + \sum^{\min(p_{2},\ell)}_{s=\max(\ell-k+1,-p_{1})}\overline{\widetilde{\bi{H}_{1}}(s)}\bi{J}^{2}_{M'}\overline{\bi{H}_{1}(\ell-s)}. \label{eq:HSUsomme}
\end{align}
Since $\bi{J}^{2}_{M'}=\bi{I}_{M'}$, the previous equation can be seen as the sum of a complex matrix with its conjugate, leading to a real matrix.
We deduce that
$$ \frac{1}{2}\delta_{\ell}\bi{I}_{N}
=\sum^{\min(p_{2},\ell)}_{s=\max(\ell-k+1,-p_{1})} \left[\widetilde{\bi{H}^{R}_{1}}(s), -\widetilde{\bi{H}^{I}_{1}}(s)\right]\left(\begin{array}{c}
\bi{H}^{R}_{1}(\ell-s)\\ 
\bi{H}^{I}_{1}(\ell-s)\\
\end{array} \right), $$
where $\bi{A}^{R}$ is the matrix of the real part of a matrix $\bi{A}$ and $\bi{A}^{I}$ is its imaginary part. 
We will then define the following matrices:
\begin{equation}\widetilde{{\cal{H}}}^\top_{s}=\left[\widetilde{\bi{H}^{R}_{1}}(-p_{1}),-\widetilde{\bi{H}^{I}_{1}}(-p_{1}),\cdots,\widetilde{\bi{H}^{R}_{1}}(p_{2}),-\widetilde{\bi{H}^{I}_{1}}(p_{2})\right]\in\RR^{N\times pM},
\label{eq:defhtildeMpair}
\end{equation}
 and 
$${\cal{H}}^\top_{s}=\left(\begin{array}{ccccc}
 \bi{H}^{R}_{1}(0) & \cdots & \bi{H}^{R}_{1}(k-1) &  &  0\\
 \bi{H}^{I}_{1}(0) & \cdots & \bi{H}^{I}_{1}(k-1) &  &  0\\
  &  \ddots & &\ddots &  \\
  0 &   & \bi{H}^{R}_{1}(0) &  \cdots & \bi{H}^{R}_{1}(k-1) \\
  0 &   & \bi{H}^{I}_{1}(0) &  \cdots & \bi{H}^{I}_{1}(k-1) \\
  \end{array}\
\right)\in\RR^{pM\times(k+p_{1}+p_{2})N}.$$

\paragraph{Second case: $M$ is odd }
\label{sssec:Mimpair}

Similarly to the first case, Conditions \eqref{eq:cond1} and \eqref{eq:cond2} can be rewritten:
\begin{align*}
(\ref{eq:cond1})&\Leftrightarrow\forall \ell\in\left\{-p_{1},...,p_{2}\right\}, \left\{ \begin{array}{l} \widetilde{\bi{H}}(\ell)=\left[\widetilde{\bi{H}_{1}}(\ell), \bb{c}_{1}(\ell), \widetilde{\bi{H}_{2}}(\ell)\right],\, \widetilde{\bi{H}_{1}}\in \CC^{N\times M'}\mbox{ and }\widetilde{\bi{H}_{2}} \in \CC^{N\times M'}, \\
\widetilde{\bi{H}_{1}}(\ell)= \overline{\widetilde{\bi{H}_{2}}(\ell)}\bi{J}_{M'} \mbox{ and } \bb{c}_{1}(\ell)\in\RR^{N}\\
\end{array}
\right.\\
(\ref{eq:cond2})&\Leftrightarrow\forall \ell\in\left\{0,...,k-1\right\}, \left\{ \begin{array}{l} \bi{H}(\ell)=\left(
\begin{array}{c}
\bi{H}_{1}(\ell)\\ 
\bb{c}_{2}(\ell)^{\top}\\
\bi{H}_{2}(\ell)\\
\end{array}
\right),\, \bi{H}_{1}\in \CC^{M'\times N}\mbox{ and }\bi{H}_{2} \in \CC^{M'\times N}, \\
\bi{H}_{1}(\ell)= \bi{J}_{M'}\overline{\bi{H}_{2}(\ell)}  \mbox{ and } \bb{c}_{2}(\ell)\in\RR^{N}\\
\end{array}
\right.
\end{align*}
with $M'=(M-1)/2$. 
Combining these conditions  with the PR equation and following the same reasoning as in the previous section, we deduce that
$$ \frac{1}{2}\delta_{\ell}\bi{I}_{N}=\sum^{\min(p_{2},\ell)}_{s=\max(\ell-k+1,-p_{1})} \left[\widetilde{\bi{H}^{R}_{1}}(s),\frac{\bb{c}_{1}(s) }{\sqrt{2}}, -\widetilde{\bi{H}^{I}_{1}}(s)\right]\left(\begin{array}{c}
\bi{H}^{R}_{1}(\ell-s)\\
\frac{\bb{c}_{2}(\ell-s)^{\top} }{\sqrt{2}}\\
\bi{H}^{I}_{1}(\ell-s)\\
\end{array} \right). $$
Subsequently, we introduce in this case the following matrices:
\begin{equation}
\widetilde{{\cal{H}}}^\top_{s}= \left[\widetilde{\bi{H}^{R}_{1}}(-p_{1}),\frac{\bb{c}_{1}(-p_{1})}{\sqrt{2}},-\widetilde{\bi{H}^{I}_{1}}(-p_{1}),\cdots,\widetilde{\bi{H}^{R}_{1}}(p_{2}),\frac{\bb{c}_{1}(p_{2})}{\sqrt{2}},-\widetilde{\bi{H}^{I}_{1}}(p_{2})\right]\in\RR^{N\times pM},
\label{eq:defhtildeMimpair}
\end{equation}
and
$${\cal{H}}^\top_{s}=\left(\begin{array}{ccccc}
 \bi{H}^{R}_{1}(0) & \cdots & \bi{H}^{R}_{1}(k-1) &  &  0\\
 \frac{\bb{c}_{2}(0)^{\top}}{\sqrt{2}} & \cdots & \frac{\bb{c}_{2}(k-1)^{\top}}{\sqrt{2}} & & 0\\
 \bi{H}^{I}_{1}(0) & \cdots & \bi{H}^{I}_{1}(k-1) &  &  0\\
  &  \ddots & &\ddots &  \\
  0 &   & \bi{H}^{R}_{1}(0) &  \cdots & \bi{H}^{R}_{1}(k-1) \\
  0 &   & \frac{\bb{c}_{2}(0)^{\top}}{\sqrt{2}} & \cdots & \frac{\bb{c}_{2}(k-1)^{\top}}{\sqrt{2}}\\
  0 &   & \bi{H}^{I}_{1}(0) &  \cdots & \bi{H}^{I}_{1}(k-1) \\
  \end{array}\
\right)\in\RR^{pM\times(k+p_{1}+p_{2})N}.$$

\paragraph{Conclusion }

In both even and odd options, we solve a linear system of the same size as the one of Section \ref{subsec:comp}, but with real coefficients in this case. 
More precisely, with the introduced notations, we have:
$${\cal{H}}_{s}\widetilde{{\cal{H}}_{s}}={\cal{U}}_{s}
= \frac{1}{2}\left[\bi{0}_{N,p_{1}N},\;\bi{I}_{N},\;\bi{0}_{N,(p_{2}+k-1)N}\right]^\top.$$
The system is then solved, in the same way as in Section \ref{subsec:comp}. For increasing values of $p$ (starting with $p=1$), for each couple $(p_{1},p_{2})\in\NN^{2}$ such that $p=p_{1}+p_{2}+1$ we try to invert the generated system through a Moore-Penrose pseudo-inversion.

\section{Optimization}
\label{sec:tspoptim}

\subsection{Dimension reduction}
\label{sec:dimred}

\subsubsection{General case}
\label{subsec:dimredgen}
Before addressing the issue of optimization in itself, let us rewrite the linear system expressing the PR property. 
The analysis FB is still supposed invertible. Let $r$ be the rank of the matrix $\mathcal{H}\in\CC^{(k+p_{1}+p_{2})N\times pM}$. We assumed that $r< Mp$ (with $p=p_{1}+p_{2}+1$). Performing a Singular Value Decomposition \cite{Strang_G_1998_book_int_la} (SVD)  on this matrix yields 
$
\mathcal{H} = \mathcal{U}_0 \Sigma_0 \mathcal{V}_0^*\,,
$ 
where $\Sigma_0 \in \CC^{r\times r}$ is an invertible diagonal matrix, $\mathcal{U}_0 \in \CC^{N(k+p-1)\times r}$
and $ \mathcal{V}_0 \in \CC^{Mp \times r}$ are semi-unitary matrices (i.e. $\mathcal{U}^{*}_0\mathcal{U}_0=\bi{I}_{r}$ and  $\mathcal{V}^{*}_0\mathcal{V}_0=\bi{I}_{r}$). Therefore, there exists
$\mathcal{U}_1 \in \CC^{N(k+p-1)\times (N(k+p-1)-r)}$
and $\mathcal{V}_1 \in \CC^{Mp \times (Mp-r)}$ such that 
$[\mathcal{U}_0,\;\mathcal{U}_1]$ and $[\mathcal{V}_0,\;\mathcal{V}_1]$ are unitary matrices. When an inverse polyphase transfer matrix exists, a particular solution to \eqref{eq:HtildeHU} is
$\widetilde{\mathcal{H}}^0 = \mathcal{H}^\sharp \mathcal{U}\,,$
where $\mathcal{H}^\sharp=\mathcal{V}_0 \Sigma_0^{-1} \mathcal{U}_0^*$ is the pseudo-inverse matrix of $\mathcal{H}$. Equation \eqref{eq:HtildeHU} is then equivalent to 
$ 
\mathcal{U}_0 \Sigma_0 \mathcal{V}_0^*
(\widetilde{\mathcal{H}}-\widetilde{\mathcal{H}}^0) = 
\bi{0}_{N(k+p-1)\times N}.
$ 
Since $\mathcal{U}_0^*\mathcal{U}_0 = \bi{I}_r$ and $\Sigma_0$ is invertible, we get:
$\mathcal{V}_0^*(\widetilde{\mathcal{H}}-\widetilde{\mathcal{H}}^0) = \bi{0}_{r\times N}$. In other words,  the columns of  $\widetilde{\mathcal{H}}-\widetilde{\mathcal{H}}^0$ belong to $\mathrm{Ker}(\mathcal{V}_0^*)$, the null space of $\mathcal{V}_0^*$. Moreover, it can be easily seen that $\mathrm{Ker}(\mathcal{V}_0^*)$ is equal to $\mathrm{Im}(\mathcal{V}_1)$. We then obtain the following affine form for $\mathcal{H}$:
\begin{equation}
\widetilde{\mathcal{H}} = \mathcal{V}_1 \mathcal{C}+\widetilde{\mathcal{H}}^0,
\label{eq:afftildeH}
\end{equation}
where $\mathcal{C} \in \CC^{(Mp-r)\times N}$. 

The construction of a synthesis FB  thus amounts to the choice of $\mathcal{C}$. If $\mathcal{C} = \bi{0}_{(Mp-r)\times N}$, then the obtained synthesis FB is the PI FB. This expression can be further rewritten into a more convenient form for optimization purposes. First, we define the matrices $(\bi{V}_{j})_{j\in \{0,\ldots,M-1\}}$ by: for all $\ell \in \{-p_{1},\ldots,p_{2}\}$ and $n \in \{0,\ldots,Mp-r-1\}$,
$
(\bi{V}_{j})_{\ell+p_{1},n} =\mathcal{V}_1\big((\ell+p_{1})M+j,n\big),
$ 
with $\mathcal{V}_1 = [\mathcal{V}_1(s,n)]_{0\le s < Mp,0\le n <Mp-r}$.
According to \eqref{eq:afftildeH} and \eqref{eq:defHtildecur}, we  can write for all $\ell\in \left\{-p_{1},...,p_{2}\right\}$, $i\in \left\{0,...,N-1\right\}$ and $j\in \left\{0,...,M-1\right\}$:
$$
\widetilde{H}_{i,j}(\ell) = 
\sum_{n=0}^{Mp-r-1} (\bi{V}_{j})_{\ell+p_{1},n}
\mathcal{C}(n,i)+\widetilde{H}^0_{i,j}(\ell),
$$
where $\big(\widetilde{H}_{i,j}(\ell)\big)_{-p_{1}\le\ell\le p_{2}}$ represent the impulse responses of the synthesis FB, $\big(\widetilde{H}^{0}_{i,j}(\ell)\big)_{-p_{1}\le\ell\le p_{2}}$ correspond to the PI solution and $\mathcal{C} = [\mathcal{C}(n,i)]_{0\le n <Mp-r,0 \le i < N}$. For all $j\in \left\{0,..., M-1\right\}$, we introduce the matrices $\widetilde{\bi{H}}^{0}_{j}$ defined by: 
$\left(\widetilde{\bi{H}}^{0}_{j}\right)_{\ell+p_{1},i}=\widetilde{H}^0_{i,j}(\ell)$ for all $\ell\in \left\{-p_{1},..., p_{2}\right\}$ and $i\in \left\{0,..., N-1\right\}$. We thus obtain
\begin{equation}
\label{eq:redform}
\widetilde{H}_{i,j}(\ell)=\left(\bi{V}_{j}\mathcal{C}+\widetilde{\bi{H}}^{0}_{j}\right)_{\ell+p_{1},i}. 
\end{equation}
This equation is used in Section \ref{sssec:tspgencf} to simplify the optimization problem raised by the design of the synthesis FB.

The above expressions are given in the complex case,  but they naturally remain  valid in the real case. This will be illustrated by the first example of Section \ref{sssec:resoptimgal}.

\subsubsection{Symmetric case}
\label{subsec:dimredsym}

In this section, we  adapt the results of the previous section to the HS FB case. The notations used here are similar to those introduced in Section \ref{sec:invsymcasemeth2}. 
It is worth noticing that we can calculate the matrix $\widetilde{{\cal{H}}}$ directly from $\widetilde{{\cal{H}}_{s}}$, as defined in Section \ref{sec:invsymcasemeth2} when $M$ is either even or odd, in the following way:
\begin{equation} \widetilde{{\cal{H}}}=\bi{P}_{\mbox{rc}}\widetilde{{\cal{H}}_{s}}\label{eq:HHS}
\end{equation}
where the matrix $\bi{P}_{\mbox{rc}}\in\CC^{pM\times pM}$ is the block diagonal matrix built with the block:
$$\left(\begin{array}{cc}
  \bi{I}_{M'} & -\imath\bi{I}_{M'} \\
  \bi{J}_{M'} & \imath \bi{J}_{M'} \\
 \end{array}\
\right)$$
if $M=2M'$ (even case, as seen in Section \ref{sssec:Mpair}) and:
$$
\left(\begin{array}{ccc}
  \bi{I}_{M'} & 0 & -\imath\bi{I}_{M'} \\
  0 & \sqrt{2} & 0  \\
  \bi{J}_{M'} & 0 & \imath \bi{J}_{M'} \\
  \end{array}\right)
$$
if $M=2M'+1$ (odd case, as seen in Section \ref{sssec:Mimpair}). 
By applying once again an SVD on ${\cal{H}}_{s}$, and by following the same steps as in Section~\ref{subsec:dimredgen}, we end up with an equation similar to  \eqref{eq:afftildeH}:
\begin{equation}
\widetilde{\mathcal{H}_{s}} = \mathcal{V}_1 \mathcal{C}+\widetilde{\mathcal{H}_{s}}^0.
\label{eq:afftildeHs}
\end{equation}
Note that, according to the properties of the SVD, the matrix $\mathcal{C}$ is now real-valued. By noticing that $\widetilde{\mathcal{H}}^0= \bi{P}_{\mbox{rc}}\widetilde{\mathcal{H}_{s}}^0$ and setting $\mathcal{W}_1=\bi{P}_{\mbox{rc}}\mathcal{V}_1$, we finally obtain:
\begin{equation} \widetilde{{\cal{H}}} = \mathcal{W}_1 \mathcal{C}+\widetilde{\mathcal{H}}^0. 
\label{eq:afftildeHs2}
\end{equation}
We next define the matrices $(\bi{W}_{j})_{0 \le j \le M-1}$: for all $\ell \in \{-p_{1},\ldots,p_{2}\}$ and $n \in \{0,\ldots,Mp-r-1\}$: 
$(\bi{W}_{j})_{\ell+p_{1},n} =\mathcal{W}_1\big((\ell+p_{1})M+j,n\big).$ 
Using \eqref{eq:afftildeHs2} as in Section ~\ref{subsec:dimredgen}, we get:
\begin{equation}
\label{eq:redform2}
\widetilde{H}_{i,j}(\ell)=\left(\bi{W}_{j}\mathcal{C}+\widetilde{\bi{H}}^{0}_{j}\right)_{\ell+p_{1},i}. 
\end{equation}

\subsection{Optimal solution}
\label{sec:optsol}

\subsubsection{General form for the cost functions}
\label{sssec:tspgencf}

Depending on the  desired properties for the synthesis FB, several cost functions can be employed. We first propose a generic   cost function formulation and then provide  practical examples based on the filter time or frequency spread, respectively.

Our goal is to optimize the filter shape given by the coefficients $\widetilde{h}$ of the synthesis FB, subject to the perfect reconstruction property. According to the results in Section~\ref{sec:dimred}, it is possible to represent the coefficients in the general case by using \eqref{eq:redform}. The optimization favorably takes place in the reduced dimension space the matrix $\mathcal{C}$ belongs to (compared with the dimension of the space of the coefficients of $\widetilde{h}$), thus allowing us to reformulate the optimization under a perfect reconstruction constraint as an unconstrained problem. 
In this context, the generic  cost function form we consider is:
$$
J(\widetilde{h})= \widetilde{J}(\mathcal{C})=\sum^{M-1}_{j=0}\frac{\left\| \bi{V}_j\mathcal{C}+\widetilde{\bi{H}}^{0}_{j} \right\|^{2}_{K_j}}{\left\| \bi{V}_j\mathcal{C}+\widetilde{\bi{H}}^{0}_{j} \right\|_{\Lambda}^{2}}.
$$
Hereabove, the following notation has been employed:
$$\forall\bi{A}\in\CC^{N\times p},\quad\left\|\bi{A}\right\|^{2}_{K}=\sum_{(i,i',\ell,\ell')} \bi{A}_{i,\ell}\overline{\bi{A}_{i',\ell'}}K(i,i',\ell,\ell'),$$ 
where $K$ and $\Lambda$ are $(N\times N\times p\times p)$ kernels. Moreover, we assume here that $\left\|\bi{A}\right\|_{K}$ represents a semi-norm over $\CC^{N\times p}$ and  it is thus real non-negative. Let $\bi{K}'$ be the matrice defined by $\bi{K}'_{i+\ell N,i'+\ell' N}=K(i,i',\ell,\ell')$ for all $(\ell,\ell')\in \left\{0,...,p-1\right\}^{2}$ and $(i, i')\in\left\{0,..., N-1\right\}^{2}$. Without loss of generality, this matrix can be taken positive semi-definite, which implies that
$\bi{K}'_{i+\ell N,i'+\ell' N}=\overline{\bi{K}'_{i'+\ell' N,i+\ell N}}$ and, thus,  $K(i,i',\ell,\ell')=\overline{K(i',i,\ell',\ell)}.$ We deduce the following expression:
$$
\left\|\bi{A}\right\|^{2}_{K}=\sum_{(i,i',\ell,\ell')} \bi{A}_{i,\ell}\overline{\bi{A}_{i',\ell'}}K(i,i',\ell,\ell')
= \sum_{(i,i',\ell,\ell')} \bi{A}_{i,\ell}\overline{\bi{A}_{i',\ell'}K(i',i,\ell',\ell)}.
$$
This relation will be used to simplify some equations in Section \ref{ssec:algograd} and Appendix \ref{ssec:gradfcnsym}. We finally notice that $\bi{K}'$ is a positive definite Hermitian matrix if and only if $\left\|.\right\|_{K}$ is a norm.

\subsubsection{Impulse responses optimization}
\label{subsec:optsolimpul}

A first objective  is to obtain impulse responses 
$(\widetilde{h}_{j})_{0\le j < M}$ for the synthesis filters well-localized, around some time-indices $(\overline{m}_j)_{0\le j < M}$. We now explain the link between the cost function form introduced in the previous section and the previously described dimension reduction to further simplify the problem. 

The considered cost function is the following:
$$J_{\textsf{t}}(\widetilde{h})=\sum^{M-1}_{j=0} \omega_{\textsf{t},j} \frac{\sum_{m}\left|m-\overline{m}_j\right|^{\alpha}\left|\widetilde{h}_{j}(m)\right|^{2}} {\sum_{m}\left|\widetilde{h}_{j}(m)\right|^{2}}, $$
with $\alpha\in\RR_{+}^{*}$ and weights $(\omega_{\textsf{t},j})_{0 \le j < M} \in (\mathbb{R}_+)^M$
such that $\sum_{j=1}^{M-1} \omega_{\textsf{t},j} = 1$. \\
If  $\alpha=2$ and $\overline{m}_j=\frac{\sum_{m}m\left|\widetilde{h}_{j}(m)\right|^{2}}{\sum_{m}\left|\widetilde{h}_{j}(m)\right|^{2}}$, then  $J_{\textsf{t}}(\widetilde{h})$ represents a weighted sum of the standard temporal dispersions measuring the time localization of a filter $\widetilde{h}_{j}$ \cite{Flandrin_P_1998_book_tim_tfanalysis}.
Combined with  \eqref{eqhtildeimpul}, we get:
\begin{align*}
J_{\textsf{t}}(\widetilde{h})=&\sum^{M-1}_{j=0} \omega_{\textsf{t},j} \frac{\sum^{p_{2}}_{\ell=-p_{1}}\sum^{N-1}_{i=0} \left|\ell N-i-\overline{m}_j\right|^{\alpha}\left|\widetilde{h}_{j}(\ell N-i)\right|^{2}} {\sum^{p_{2}}_{\ell=-p_{1}}\sum^{N-1}_{i=0}\left|\widetilde{h}_{j}(\ell N-i)\right|^{2}} \\
=&\sum^{M-1}_{j=0} \omega_{\textsf{t},j} \frac{\sum^{p_{2}}_{\ell=-p_{1}}\sum^{N-1}_{i=0}\left|\ell N-i- \overline{m}_j\right|^{\alpha}\left|\widetilde{H}_{i,j}(\ell)\right|^{2}} {\sum^{p_{2}}_{\ell=-p_{1}}\sum^{N-1}_{i=0}\left|\widetilde{H}_{i,j}(\ell)\right|^{2}}.
\end{align*}
We now introduce the kernels $K^{\textsf{t}}_{j}$ and $\Lambda$ defined by 
\begin{align}
K^{\textsf{t}}_{j}(i,i',\ell+p_{1},\ell'+p_{1})&=\omega_{\textsf{t},j} \left|\ell N-i- \overline{m}_j\right|^{\alpha}\delta_{i-i'}\delta_{\ell-\ell'}, \label{eq:noyautempdef}\\
\Lambda(i,i',\ell+p_{1},\ell'+p_{1})&=\delta_{i-i'}\delta_{\ell-\ell'}, \notag
\end{align}
for all $j\in \{0,\ldots,M-1\}$, $(\ell,\ell')\in \{-p_1,\ldots,p_2\}^2$
and $(i,i') \in \{0,\ldots,N-1\}^2$.
Using \eqref{eq:redform} we write:  
$$\sum^{p_{2}}_{\ell=-p_{1}}\sum^{N-1}_{i=0}\left|\widetilde{H}_{i,j}(\ell)\right|^{2}
=\left\| \bi{V}_{j}\mathcal{C}+\widetilde{\bi{H}}^{0}_{j} \right\|^2_{\Lambda} $$
and  
$$\omega_{\textsf{t},j} \sum^{p_{2}}_{\ell=-p_{1}}\sum^{N-1}_{i=0}(\ell N-i- \overline{m}_j)^2\left|\widetilde{H}_{i,j}(\ell)\right|^{2}
=\left\| \bi{V}_{j}\mathcal{C}+\widetilde{\bi{H}}^{0}_{j} \right\|^{2}_{K^{\textsf{t}}_{j}}. $$
Hereabove, $\left\|.\right\|_{\Lambda}$ reduces to the Frobenius norm.
Finally, we deduce that
$$
J_{\textsf{t}}(\widetilde{h})=\sum^{M-1}_{j=0}\frac{\left\| \bi{V}_j\mathcal{C}+\widetilde{\bi{H}}^{0}_{j} \right\|^{2}_{K^{\textsf{t}}_j}}{\left\| \bi{V}_j\mathcal{C}+\widetilde{\bi{H}}^{0}_{j} \right\|^{2}_{\Lambda}}= \widetilde{J_{\textsf{t}}}(\mathcal{C}).
$$
The constrained minimization of $J_{\textsf{t}}$ is then reexpressed as the unconstrained minimization of $\widetilde{J_{\textsf{t}}}$.

\subsubsection{Frequency response optimization}
\label{subsec:optsolfreq}

We proceed similarly to the previous section, for a different cost function $J_{\textsf{f}}(\widetilde{h})$. Our goal, dual to that in the previous section, is now to regularize the frequency responses of the synthesis FB by concentrating the frequency response of each filter $\widetilde{h}_{j}$ around some frequency $f_j$. This is achieved by minimizing
\begin{equation}
\label{eq:costF}
J_{\textsf{f}}(\widetilde{h})=\sum^{M-1}_{j=0} 
\omega_{\textsf{f},j}
\frac{\int^{1/2+f_j}_{-1/2+f_j}\left|\nu-f_j\right|^\alpha\left|\widetilde{h}_{j}[\nu]\right|^{2}d\nu} {\int^{1/2+f_j}_{-1/2+f_j}\left|\widetilde{h}_{j}[\nu]\right|^{2}d\nu}
\end{equation}
where $\alpha\in\RR_{+}^{*}$, $(\omega_{\textsf{f},j})_{0 \le j < M} \in
(\RR_+^*)^M$ with $\sum_{j=0}^{M-1}\omega_{\textsf{f},j} = 1$ 
 and, $\widetilde{h}_{j}[.]$ is the frequency response of the $j^{\mbox{th}}$ synthesis filter, defined as
$$\forall \nu\in [-1/2,1/2[,\quad
\widetilde{h}_{j}[\nu]=\sum^{p_{2}}_{\ell=-p_{1}}\sum^{N-1}_{i=0}\widetilde{H}_{i,j}(\ell)e^{-2\imath\pi (N\ell-i)\nu}.$$
When $\displaystyle{f_{j}=\frac{\int^{1/2}_{-1/2}\nu\left|\widetilde{h}_{j}[\nu]\right|^{2}d\nu} {\int^{1/2}_{-1/2}\left|\widetilde{h}_{j}[\nu]\right|^{2}d\nu}}$, the cost function $J_{\textsf{f}}(\widetilde{h})$ represents a classical weighted frequency dispersion measure for the synthesis filters. 
We then define the kernel 
\begin{align}
K^{\textsf{f}}_{j}(i,i',\ell+p_{1},\ell'+p_{1})&=
\omega_{\textsf{f},j}
\int^{1/2+f_j}_{-1/2+f_j}\left|\nu-f_{j}\right|^{\alpha}e^{-2\imath\pi (N(\ell-\ell')-(i-i'))\nu} d\nu \notag \\
&=\omega_{\textsf{f},j}\int^{1/2}_{-1/2}\left|\nu\right|^{\alpha}e^{-2\imath\pi (N(\ell-\ell')-(i-i'))(\nu+f_{j})} d\nu,\label{eq:noyaufreqdef}
\end{align}
with $(i,i',\ell,\ell')\in \left\{0,...,N-1\right\}^{2}\times\left\{-p_{1},...,p_{2}\right\}^{2}$.

\begin{remark}
The examples provided in Section~\ref{sssec:tspoptfbex} are obtained with $\alpha=2$. In this case, the explicit expression of the kernel becomes:
$$K^{\textsf{f}}_{j}(i,i',\ell+p_{1},\ell'+p_{1}) =  \begin{cases}
\displaystyle \frac{\omega_{\textsf{f},j}}{12} & \mbox{if $i=i'$ and $\ell=\ell'$}\\
\displaystyle \frac{\omega_{\textsf{f},j}(-1)^{N(\ell-\ell')-(i-i')}e^{-2\imath\pi(N(\ell-\ell')-(i-i'))f_j}}{2\pi^2(N(\ell-\ell')-(i-i'))^2} & \mbox{otherwise}
\end{cases}$$
with $(i,i',\ell,\ell')\in \left\{0,...,N-1\right\}^{2}\times\left\{-p_{1},...,p_{2}\right\}^{2}$.
\end{remark}

Combining these notations and \eqref{eq:redform}, we have
$$\omega_{\textsf{f},j}\int^{1/2+f_j}_{-1/2+f_j}\left|\nu-f_j\right|^{\alpha}\left|\widetilde{h}_{j}[\nu]\right|^{2}d\nu=\left\|\bi{V}_{j}\mathcal{C}+\widetilde{\bi{H}}^{0}_{j}\right\|^{2}_{K^{\textsf{f}}_{j}}. $$
Invoking Plancherel's theorem and the kernel $\Lambda$ defined in Section~\ref{subsec:optsolimpul}, we obtain:
$$\int^{1/2+f_j}_{-1/2+f_j}\left|\widetilde{h}_{j}[\nu]\right|^{2}d\nu =\sum^{p_{2}}_{\ell=-p_{1}}\sum^{N-1}_{i=0}\left|\widetilde{H}_{i,j}(\ell)\right|^2
= \left\|\bi{V}_{j}\mathcal{C}+\widetilde{\bi{H}}^{0}_{j}\right\|^{2}_{\Lambda}.$$
Finally, substituting these expressions in \eqref{eq:costF} yields
\begin{align*}
J_{\textsf{f}}(\widetilde{h}) &= \sum^{M-1}_{j=0}\frac{\left\|\bi{V}_{j}\mathcal{C}+\widetilde{\bi{H}}^{0}_{j}\right\|^{2}_{K^{\textsf{f}}_{j}}}{\left\|\bi{V}_{j}\mathcal{C}+\widetilde{\bi{H}}^{0}_{j}\right\|^{2}_{\Lambda}}
= \widetilde{J_{\textsf{f}}}(\mathcal{C}).
\end{align*}
Once again, the constrained optimization problem has been reformulated as an unconstrained  one.

\subsection{Gradient optimization}
\label{ssec:algograd}

The constrained optimization problem being turned  into an unconstrained minimization, we now provide more details about the minimization algorithm we employ. In this work we have used a simple gradient algorithm with an adaptive step $\mu_n$. 
The algorithm can be summarized as follows:
\begin{dingautolist}{172}
	\item Initialization: $\mathcal{C}_{0}=\bi{0}$, $n=0$.
	\item $\mu_n=1$
	\item Computation of $\bi{D}_{n}=\nabla \widetilde{J}(\mathcal{C}_{n})$. 
	\item While $\widetilde{J}(\mathcal{C}_{n}-\mu_n\bi{D}_{n})\ge\widetilde{J}(\mathcal{C}_{n})$, set $\mu_n\gets\frac{1}{\frac{1}{\mu_n}+1}$.
	\item $\mathcal{C}_{n+1}=\mathcal{C}_{n}-\mu_n\bi{D}_{n}$.
  \item If $\left\|\mathcal{C}_{n+1}-\mathcal{C}_{n} \right\|>\epsilon$ then 
increment $n$ and go to step \ding{173}.
\end{dingautolist}

The step-size $\mu_n$ used here remains large as long as the algorithm is getting closer to a local minimum (in other words, as long as $\widetilde{J}(\mathcal{C}_{n+1})<\widetilde{J}(\mathcal{C}_{n})$). It is only adapted (reduced) to prevent  the criterion from increasing. The initialization with $\mathcal{C}_{0}=\bi{0}$ entails that we consider the pseudo-inverse synthesis FB as the starting point for the algorithm. In practice, $\epsilon$ was set to $10^{-13}$.

Other step selection strategies exist: constant or optimal steps, steps satisfying Wolfe or Armijo conditions \cite{Nocedal_J_1999_book_num_o,Barrault_M_1999_tr_opt_ndapi}. 
The method used in this work is easy to implement and is well-suited to the different cost functions we have considered, while keeping a reasonable complexity.

As the cost functions considered in this work are not convex, there is no theoretical guarantee that the algorithm converges to a global minimum. Yet, as is shown in Section~\ref{sec:tspexamples}, initializing this method with the PI synthesis FB provides quite good results and extensive simulations have confirmed this good behaviour.

The expression of the gradient for the general cost function is given in Appendix~\ref{subsec:gradgal} and is next applied to $\widetilde{J_{\textsf{t}}}$ and $\widetilde{J_{\textsf{f}}}$ in Appendix~\ref{subsec:genex}.

\subsection{Optimal solution in the symmetric case}

\subsubsection{Cost functions}

Using the same notations as in Section~\ref{sec:optsol}, in the HS case, the following form of the cost function is found:
$$
J_{\textsf{s}}(\widetilde{h})= \widetilde{J_{\textsf{s}}}(\mathcal{C})=\sum^{M-1}_{j=0}\frac{\left\| \bi{W}_j\mathcal{C}+\widetilde{\bi{H}}^{0}_{j} \right\|^{2}_{K_j}}{\left\| \bi{W}_j\mathcal{C}+\widetilde{\bi{H}}^{0}_{j} \right\|_{\Lambda}^{2}}.
$$
As in the general case, \eqref{eq:redform2} has been used to transform the constrained optimization problem on $\widetilde{h}$ into an unconstrained minimization problem on $\mathcal{C}$.

\subsubsection{Examples of cost functions}

Equation~\eqref{eq:redform2} is very similar to \eqref{eq:redform}. We consequently define the cost functions in the HS case following the same approach as in Sections \ref{subsec:optsolimpul} and \ref{subsec:optsolfreq}. Thus, the following functions are considered:
$$
\widetilde{J_{\textsf{ts}}}(\mathcal{C})=\sum^{M-1}_{j=0}\frac{\left\| \bi{W}_j\mathcal{C}+\widetilde{\bi{H}}^{0}_{j} \right\|^{2}_{K^{\textsf{t}}_j}}{\left\| \bi{W}_j\mathcal{C}+\widetilde{\bi{H}}^{0}_{j} \right\|^{2}_{\Lambda}},
$$
to concentrate the time localization of impulse responses, and:
$$\widetilde{J_{\textsf{fs}}}(\mathcal{C})=\sum^{M-1}_{j=0}\frac{\left\|\bi{W}_{j}\mathcal{C}+\widetilde{\bi{H}}^{0}_{j}\right\|^{2}_{K^{\textsf{f}}_{j}}}{\left\|\bi{W}_{j}\mathcal{C}+\widetilde{\bi{H}}^{0}_{j}\right\|^{2}_{\Lambda}}, $$
to enhance the frequency selectivity of the filters.
Their gradients are provided in Appendix~\ref{ssec:gradfcnsym}.

\section{Examples}
\label{sec:tspexamples}

As emphasized in the introduction, a wide variety of filter banks and design choices can be made. In this section, we have chosen to work with three different examples exhibiting interesting properties and allowing us to show the benefits incurred in the proposed inversion and optimization methods.

\subsection{Considered filter banks}
\label{sec:exbdf1D}

\subsubsection{Real Lapped Transforms}
\label{subsec:trnsfglt}

The study, developed for the general complex case, remains fully applicable to the design of real filter banks. As an illustration, we first consider real lapped transforms introduced in the middle of the 90s under the name of \emph{GenLOT} (generalized linear-phase lapped orthogonal transform) \cite{Queiroz_R_1996_tsp_gen_glplot}. Those transforms generalize the DCT (Discrete Cosine Transform) and the LOT (lapped orthogonal transform). 

\begin{figure*}[htbf]
\begin{center}
\begin{tabular}{c|c}
\includegraphics[height=12cm]{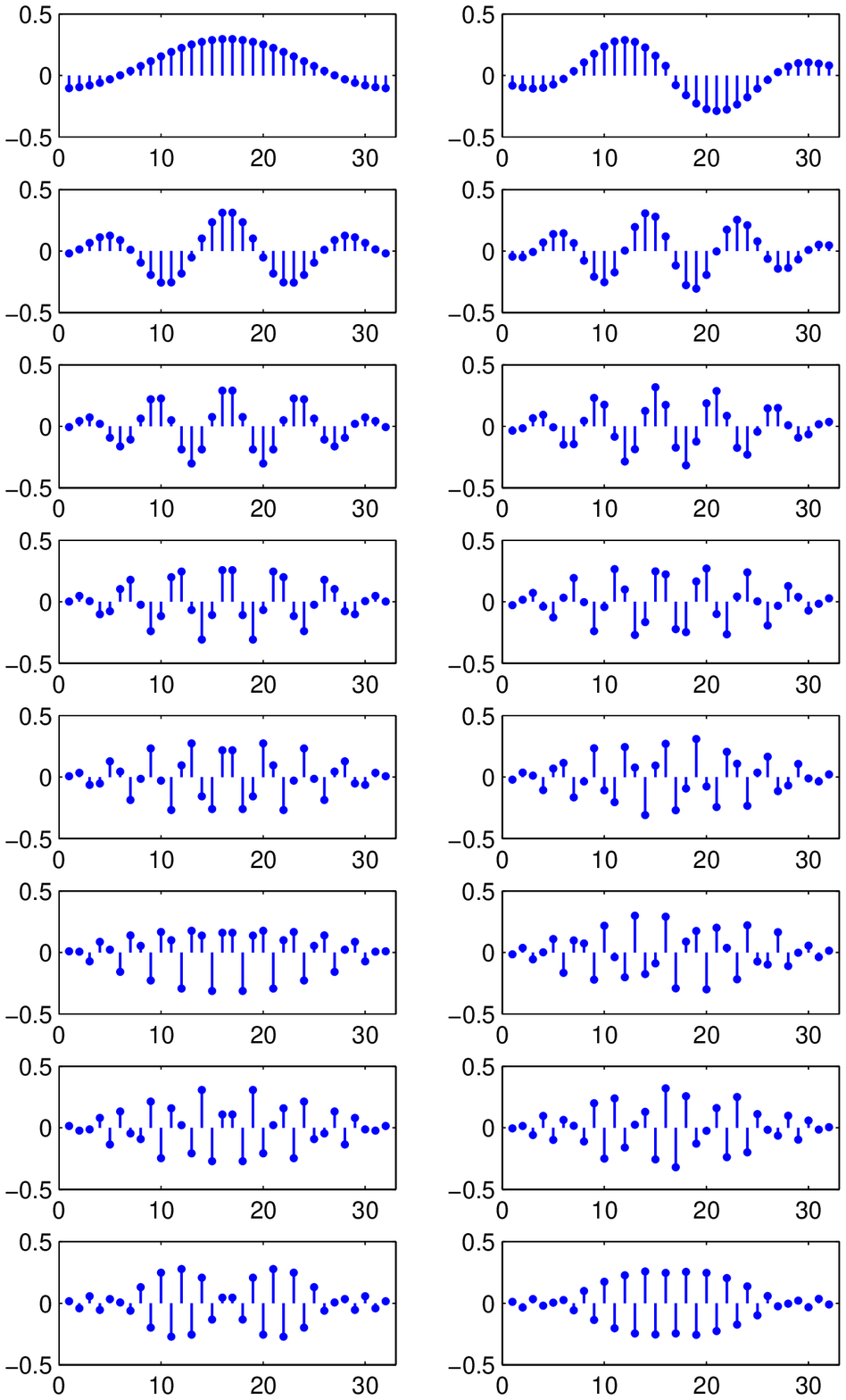} & \includegraphics[height=12cm]{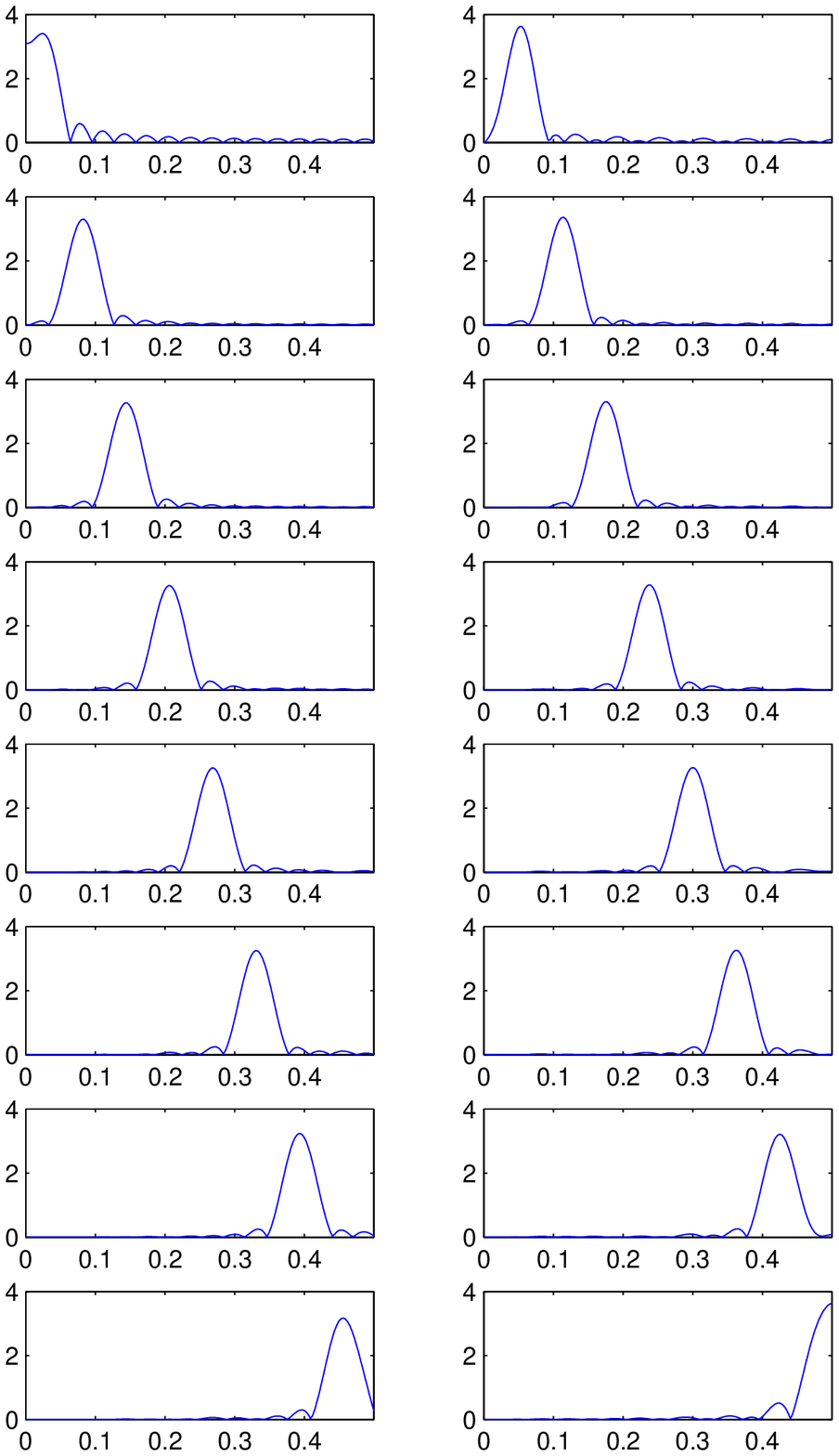}\\
(a) & (b)\\
\end{tabular}
\caption{(a) Impulse and (b) frequency responses of a GenLOT analysis FB.}
\label{fig:repfreqglt}
\end{center}
\end{figure*}

\begin{figure*}[ht]
\begin{center}
\begin{tabular}{c|c}
\includegraphics[height=12cm]{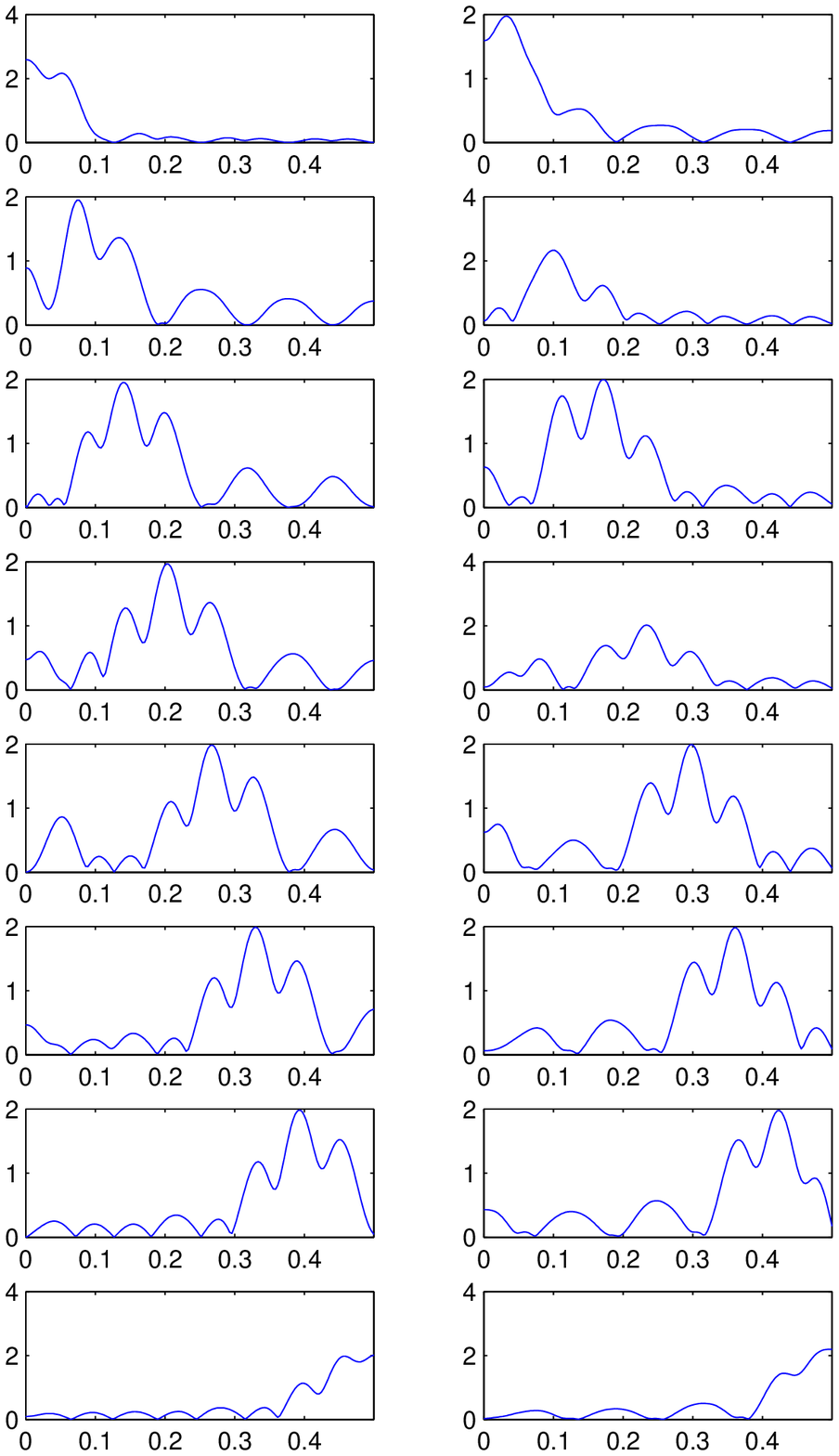}&
\includegraphics[height=12cm]{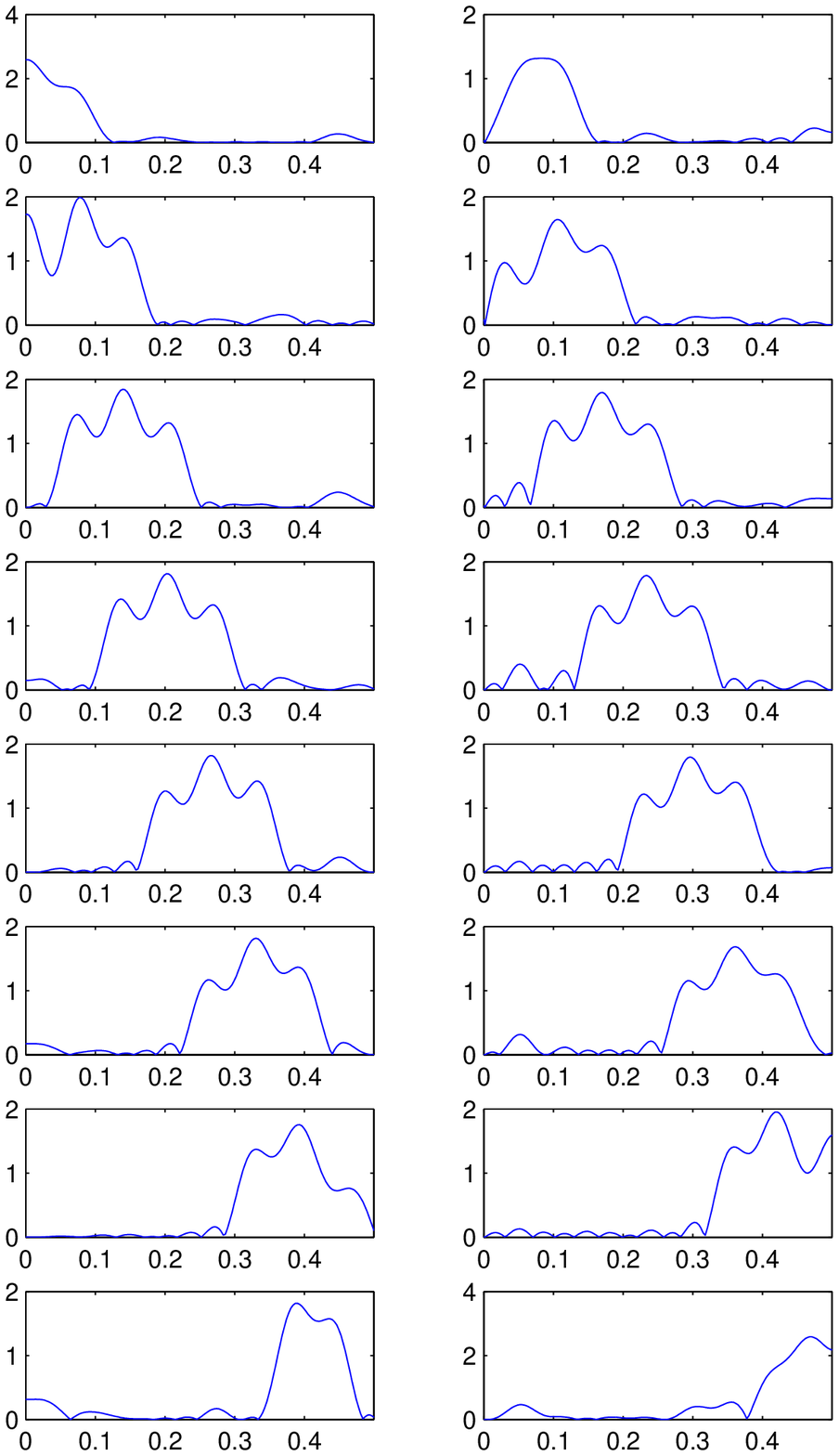}\\
(a)&(b)\\
\end{tabular}
\caption{First example (with the GenLOT FB of Section~\ref{subsec:trnsfglt}): frequency response of the synthesis FB obtained (a) through the pseudo-inverse method and (b) after optimization with cost function $\widetilde{J_{\textsf{t}}}$.}
\label{fig:repfreqgltoptim}
\end{center}
\end{figure*}

\begin{figure*}[ht]
\begin{center}
\begin{tabular}{c|c}
\includegraphics[height=12cm]{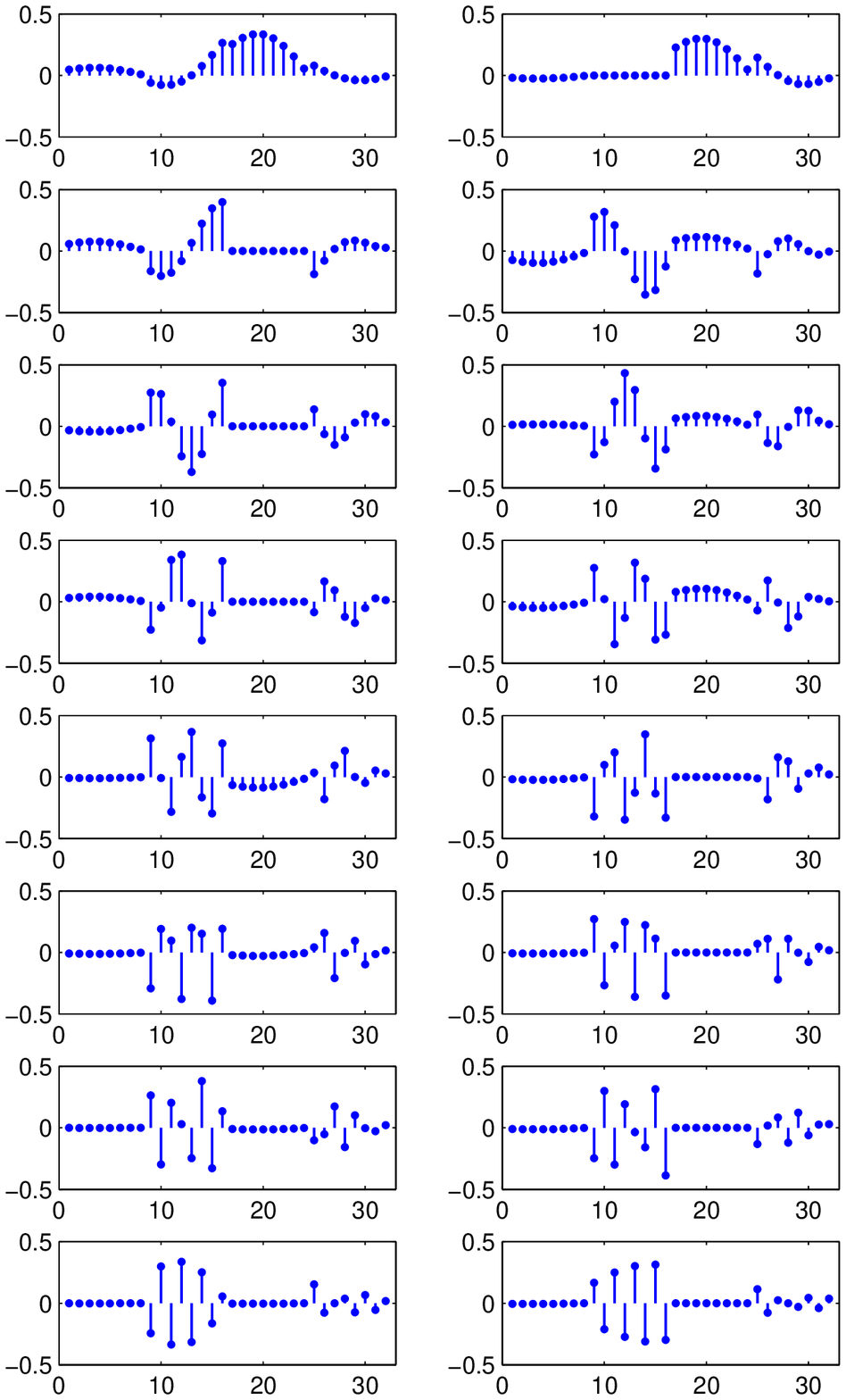}&
\includegraphics[height=12cm]{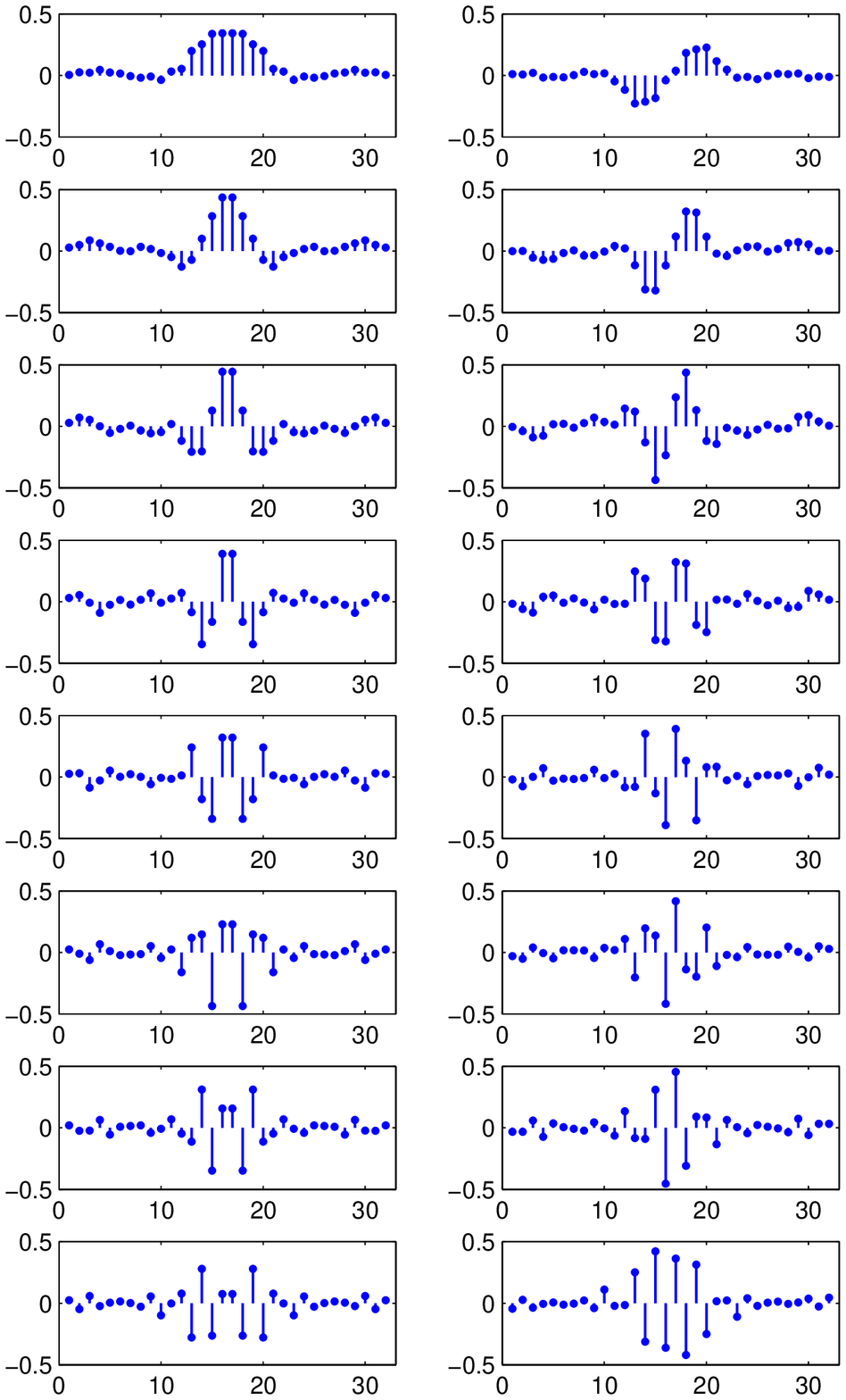}\\
(a)&(b)\\
\end{tabular}
\caption{First example: impulse response of the synthesis FB obtained (a) through the pseudo-inverse method and (b) after optimization with cost function $\widetilde{J_{\textsf{t}}}$.}
\label{fig:repimpulgltoptim}
\end{center}
\end{figure*}

To illustrate the inversion method, we have chosen a GenLOT with $M=16$ filters of $32$ coefficients. This  FB is invertible, in a non standard oversampled use, 
with parameters $N = 8$, $k = 4$ and $k'= 2$. Its impulse and frequency
responses are represented in Figure~\ref{fig:repfreqglt}. This FB is real and does not satisfy the HS condition. By using the method described in Section~\ref{subsec:comp}, we find $p_{1}=3$ and $p_{2}=0$ (hence $p=4$). The frequency and impulse responses of the synthesis FB computed with the pseudo-inverse are shown on Figures~\ref{fig:repfreqgltoptim}(a) and \ref{fig:repimpulgltoptim}(a).

\subsubsection{Modulated complex lapped transform}
\label{subsec:trnsfapp}

\begin{figure*}[ht]
\begin{center}
\begin{tabular}{c|c}
\includegraphics[height=10.3cm]{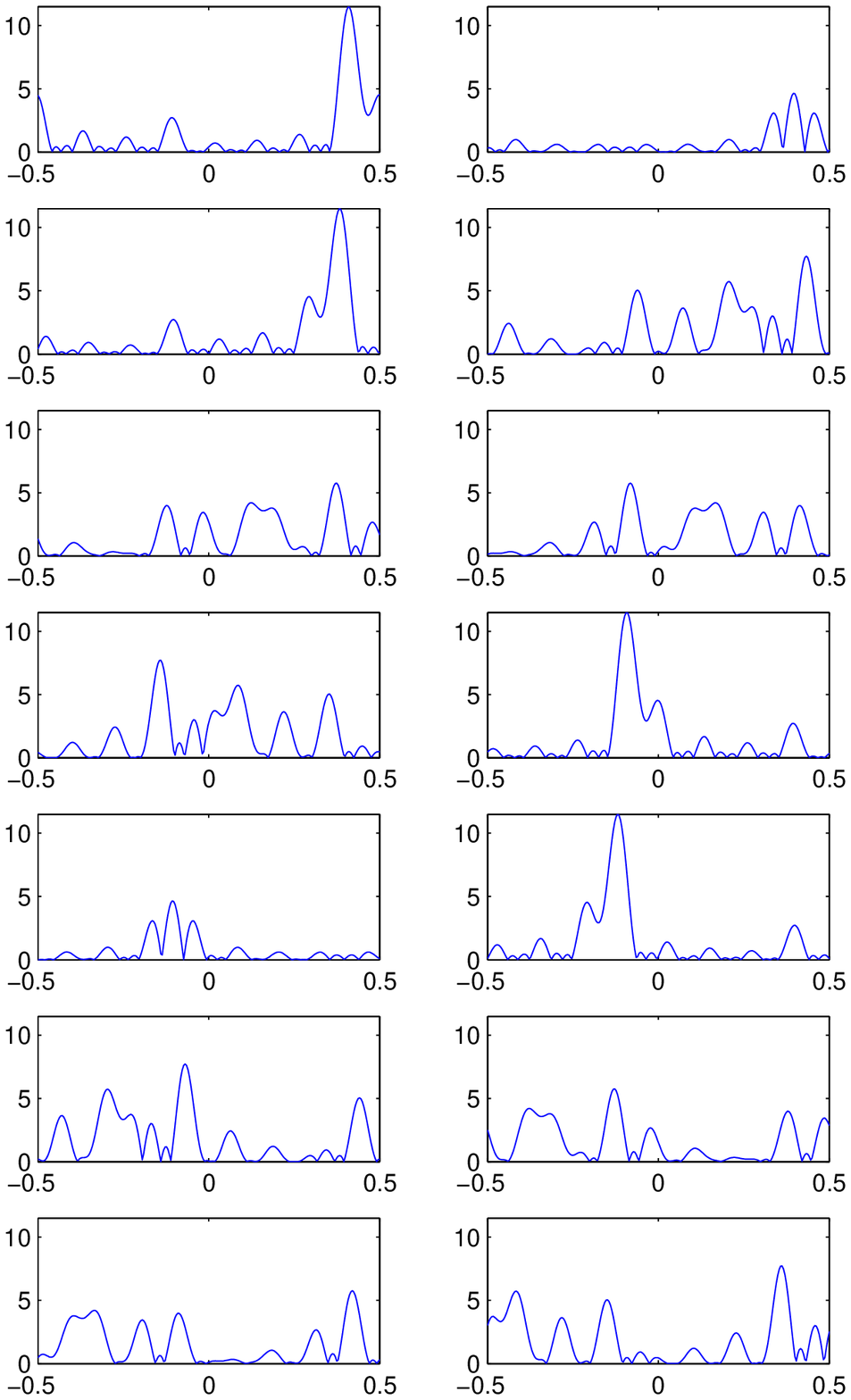}&
\includegraphics[height=10.3cm]{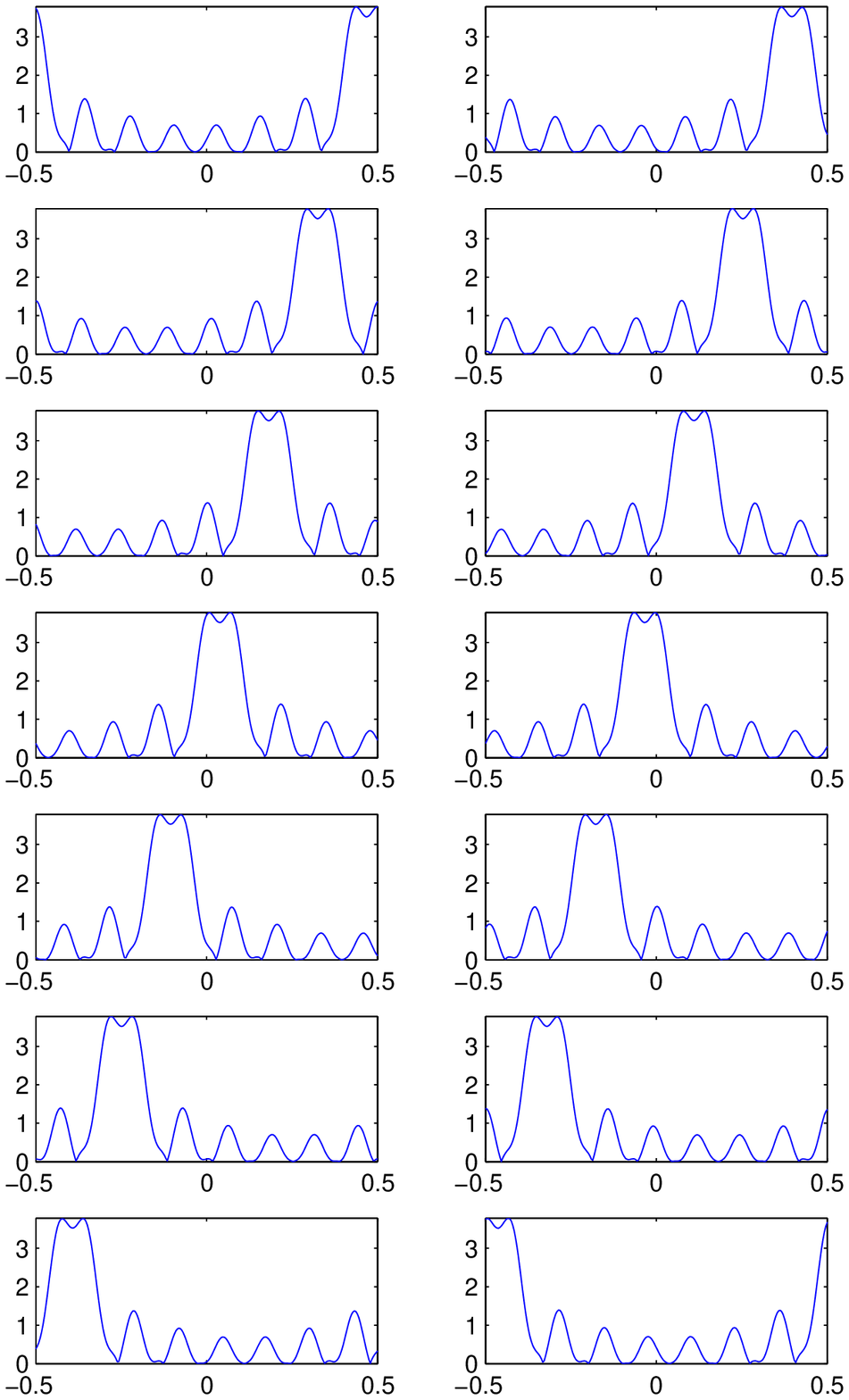}\\
(a)&(b)\\
\end{tabular}
\caption{Second example (in the MCLT case with window $h_{\textsf{a}_{1}}$): frequency responses of synthesis filters (a) before and  (b) after optimization with the cost function  $\widetilde{J_{\textsf{f}}}$.}
\label{fig:repfreqBdFoptimsin}
\end{center}
\end{figure*}

\begin{figure*}[htbf]
\begin{center}
\begin{tabular}{c|c}
\includegraphics[height=10.3cm]{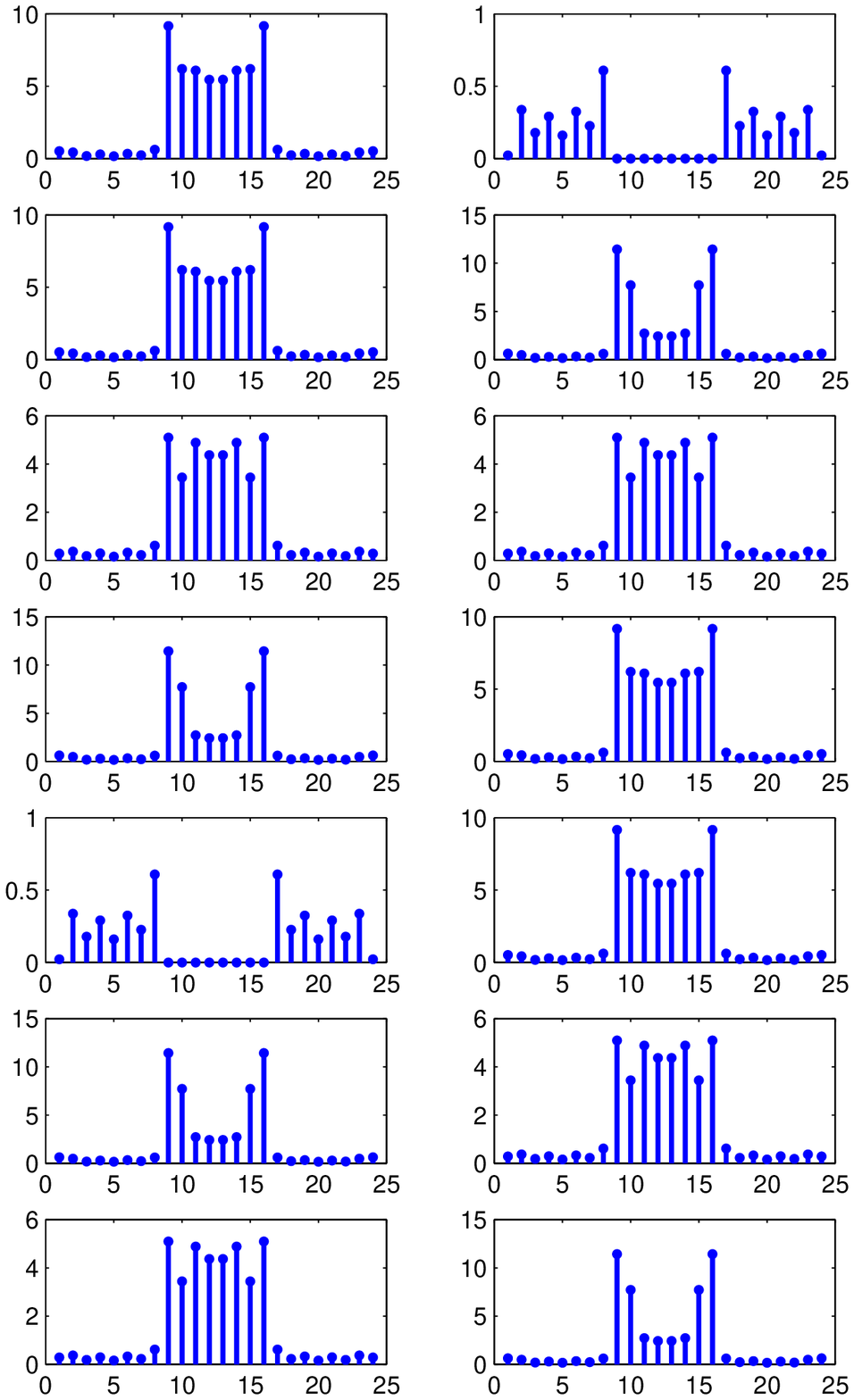}&
\includegraphics[height=10.3cm]{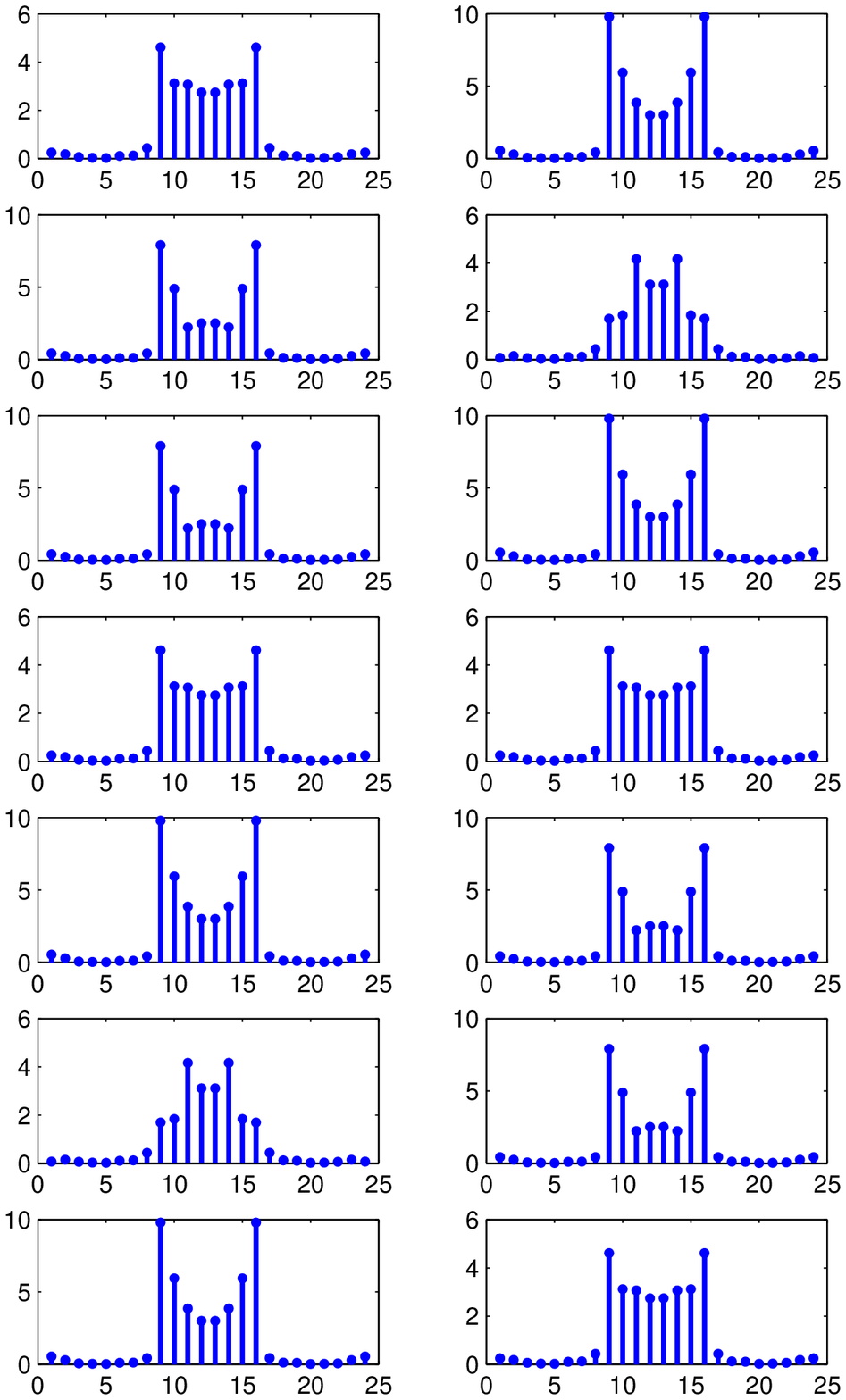}\\
(a)&(b)\\
\end{tabular}
\caption{Modulus of the impulse responses of the synthesis FBs (in the MCLT case with $h_{\textsf{a}_{2}}$ window): (a) pseudo-inverse and (b) symmetric version with the method of Section~\ref{sec:invsymcasemeth2}.}
\label{fig:repimpulBdFkaiser}
\end{center}
\end{figure*}

\begin{figure*}[htbf]
\begin{center}
\begin{tabular}{c|c}
\includegraphics[height=10.3cm]{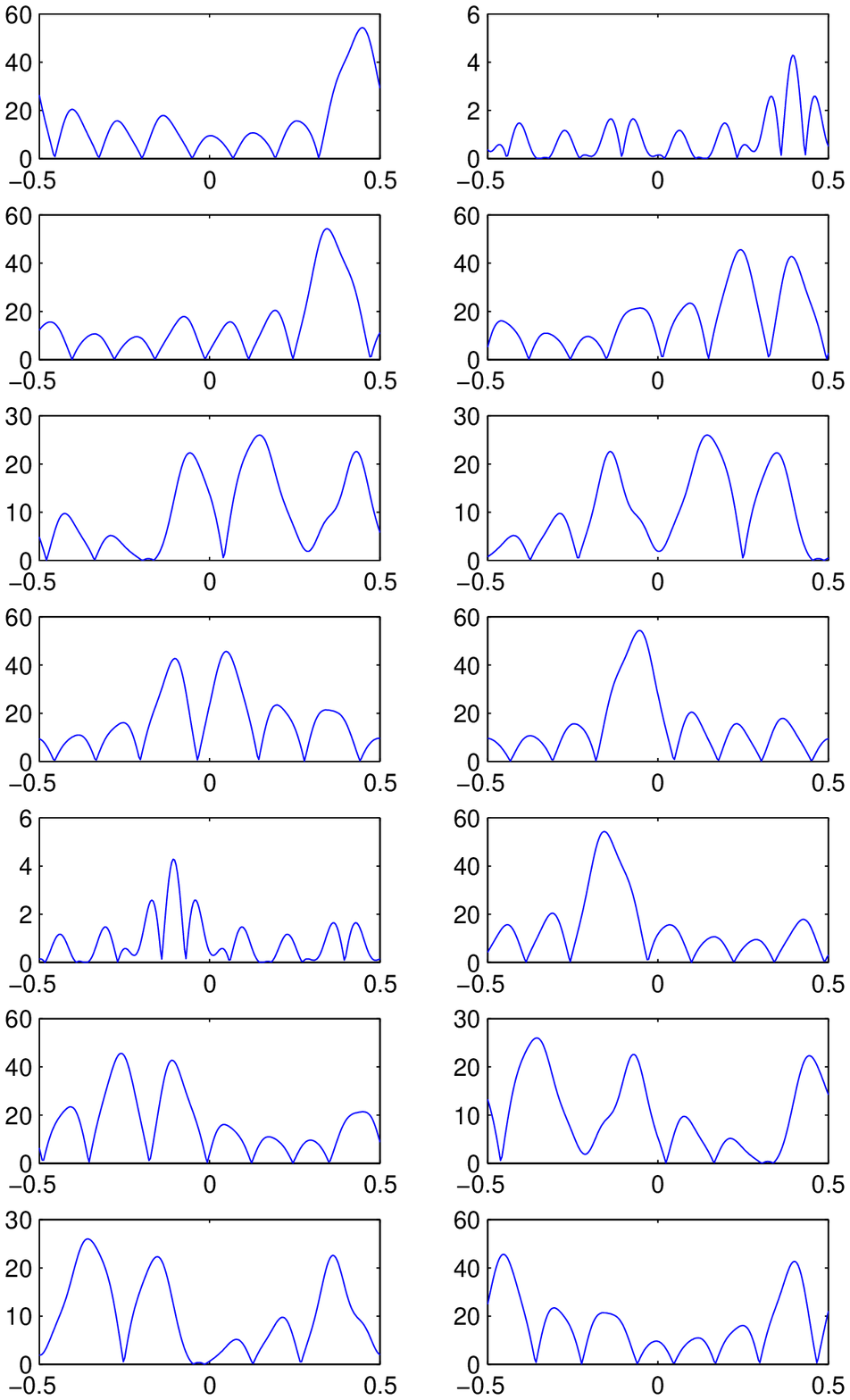}&
\includegraphics[height=10.3cm]{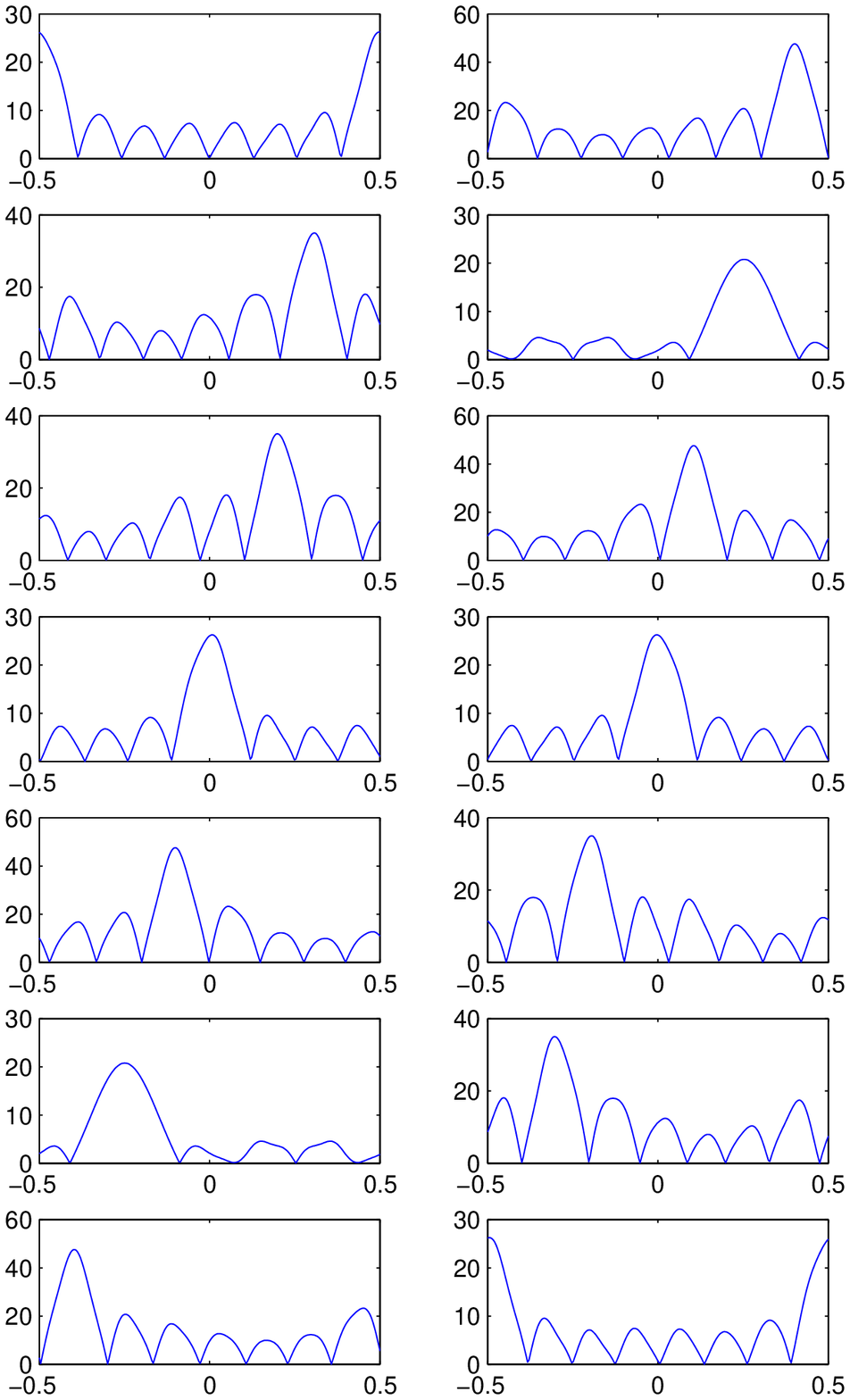}\\
(a)&(b)\\
\end{tabular}
\caption{Frequency responses of the synthesis FBs (in the MCLT case with $h_{\textsf{a}_{2}}$ window): (a) pseudo-inverse and (b) symmetric version with the method of Section~\ref{sec:invsymcasemeth2}.}
\label{fig:repfreqBdFkaiser}
\end{center}
\end{figure*}

We now consider another analysis FB based on a windowed generalized Fourier transform, corresponding to a modulated complex lapped transform (\emph{MCLT}). This family of FB has been used by Kingsbury \cite{Young_R_1993_tip_fre_dmeclt} or Malvar \cite{Malvar_H_1999_icassp_mod_cltaap} for applications in video as well as audio processing. 
The analysis impulse responses are:
$h_{i}(n)=\bi{E}(i,n)h_{\textsf{a}}(n),$ 
where
$$
\bi{E}(i,n)=\frac{1}{\sqrt{k'N}}e^{ -\imath (i-\frac{k'N}{2}+\frac{1}{2})(n-\frac{kN}{2}+\frac{1}{2})\frac{2\pi}{k'N}},
$$
and $(h_{\textsf{a}}(n))_{1\leq n \leq kN}$ is an analysis window. In this paper, we consider two analysis windows. The first, defined by 
$$\forall n\in\left\{1,...,kN\right\},\quad h_{\textsf{a}_{1}}(n)=\sin\Big(\frac{n\pi}{kN+1}\Big),$$
is a standard sine window, employed for example in \cite{Malvar_H_1999_icassp_mod_cltaap,Young_R_1993_tip_fre_dmeclt}.
The second $(h_{\textsf{a}_{2}}(n))_{1\leq n \leq kN}$, corresponds to a zero-phase low-pass filter with cutoff frequency $2\pi/(kN)$, built from a Kaiser window. This window, with better tapering than $h_{\textsf{a}_{1}}$, was used for instance in \cite{Mansour_M_2007_spl_opt_odftfb}. 
It is interesting to note that this analysis FB family, with both analysis windows, satisfies Condition~\eqref{eq:cond2}. In other words, it can be used to illustrate our approach in the HS case. 
The method from Section~\ref{sec:exist} was employed to verify the invertibility on this FB, with both analysis windows and parameters $N=8$, $k=3$ and $k'=7/4$.  
We then compute a first synthesis FB with the PI method of Section~\ref{subsec:comp}. For the analysis FB with $h_{\textsf{a}_{1}}$ window, the minimal parameters $p_{1}=2$ and $p_{2}=0$ were obtained. The frequency response of this synthesis FB is represented in Figure~\ref{fig:repfreqBdFoptimsin}(a). 
 In the $h_{\textsf{a}_{2}}$ case, the minimal parameters $p_{1}=2$ and $p_{2}=0$ were, once again, found when applying the method of Section~\ref{subsec:comp}. An HS synthesis FB is then derived from this filter bank using the method of Section~\ref{sec:invsymcasemeth2} to directly build an HS synthesis FB. 
The frequency and impulse responses of the resulting synthesis FBs, in the $h_{\textsf{a}_{2}}$ case, are shown in Figures \ref{fig:repimpulBdFkaiser} and \ref{fig:repfreqBdFkaiser}, respectively. Figure \ref{fig:repimpulBdFkaiser}(a) shows that synthesis filters present a symmetric behavior for their coefficients (in other words, they have a linear phase) while the synthesis FB in itself is not HS. We also notice that the frequency selectivity or time-frequency localization of the filters obtained through the pseudo-inverse methods is not satisfactory.

\subsection{Optimization examples}

\subsubsection{Kernel parameters}

\paragraph{Temporal kernel $K^{\textsf{t}}_{j}$}

It is defined by \eqref{eq:noyautempdef}. The parameters $\overline{m}_j$ define the (temporal) positions around which the impulse responses of the $j^{\mbox{th}}$ filter should be concentrated. To obtain well tapered filters, we need to concentrate the impulse responses around the middle of the filter support. Therefore, the same parameter was used for all filters. The support of the filters  being $\left\{-p_{1}N-N+1,..., p_{2}N\right\}$, we have chosen $\overline{m}_j$ as
$$\forall 0\le j<M,\qquad\overline{m}_j=\overline{m}=\frac{p_{2}N+1-p_{1}N-N}{2}.$$
In our design example, $\alpha$ has been set to $2$.

\paragraph{Frequency kernel $K^{\textsf{f}}_{j}$}

It is defined by \eqref{eq:noyaufreqdef}. The parameters $f_{j}$ represent the reduced frequencies around which we want to concentrate the frequency responses of the synthesis filters. More precisely, we chose $f_j$ such that it is centered inside the bandwidth of the analysis filter $h_{j}$. The exponent $\alpha$ has been set to $2$.

\paragraph{Weight parameters} 

In the proposed cost functions, the parameters $\omega_{\textsf{t},j}$ and $\omega_{\textsf{f},j}$ control the relative importance of the different filters in the optimization process. For the following examples we have chosen equal weights:
$$\forall j\in\left\{0,...,M-1\right\},\,\omega_{\textsf{t},j}=\omega_{\textsf{f},j}=\frac{1}{M}.$$
In other words, we aim at obtaining synthesis filters with similar behavior.

\subsubsection{Computation time}

In Section~\ref{sec:dimred}, we have seen how to parameterize the system and thus how to reduce the dimension of the optimization problem. To evaluate the gain resulting from this parameterization, Matlab programs were written to compare the solutions of the constrained problem, using function \verb|fmincon|, with the solutions of the unconstrained problem, using function \verb|fminunc| and using the gradient method explained in Section~\ref{ssec:algograd}. 
The two functions \verb|fmincon| and \verb|fminunc| were chosen as examples of optimization implementation, while the gradient procedure can be easily applied in different languages without requiring Matlab. These programs were tested with the analysis FB introduced in Section~\ref{subsec:trnsfapp} with the $h_{\textsf{a}_{1}}$ window and the following parameters: overlap factor $k=3$, redundancy $k'=7/4$ and downsampling $N\in\left\{ 4, 8, 16 \right\}$. The cost function used was $J_{\textsf{t}}$ (as defined in Section~\ref{subsec:optsolimpul}). 
Table~\ref{tab:tpscalculoptim} shows the computation time for the different methods on a computer with $2.16$GHz Intel Core$2$ T$7400$ CPU and $2$Gb of RAM.

\begin{table}[htbp]
\begin{center}
\caption{Computation time to optimize a synthesis FB with different methods using Matlab.}
\label{tab:tpscalculoptim}
\begin{tabular}{|l||c|c|c|}
\hline
~ & $N=4$ & $N=8$  & $N=16$ \\
\hline 
\hline
Constrained optimization (with \verb|fmincon|)  & $1.2s$ & $ 120s$ & $8800s$\\
Unconstrained optimization (with \verb|fminunc|)  & $0.04s$  & $0.7s$ & $8s$\\
Unconstrained optimization (gradient algorithm) & $0.06s$ &  $0.6s$ & $7s$ \\
\hline 
\end{tabular}
\end{center}
\end{table}

A first interesting result is that all three methods, starting from the same FB, converge to almost identical synthesis FBs. The computation times are however very different: more than two hours (with $N=16$) for the constrained optimization against a few seconds for the unconstrained optimizations. We can also notice that the gradient algorithm is as fast as the \verb|fminunc| Matlab function. This shows that the optimization method can be easily implemented, through a gradient algorithm, with no performance loss and without having to resort to the \verb|fminunc| Matlab function\footnote{This fairly sophisticated function uses an interior-reflective Newton method \cite{Coleman_T_1994_mathprg_con_irnmnmsb}.}. In other words, this last result indicates that in some applicative contexts in which Matlab is not available  the optimization method can
still be easily and efficiently implemented.

\subsubsection{Examples of optimized FBs}
\label{sssec:tspoptfbex}

In this section, we present  optimization results\footnote{A matlab toolbox for FB optimization is available here: \url{http://www.laurent-duval.eu/misc-research-codes.html}} obtained with the different FBs introduced in Section~\ref{sec:exbdf1D} and using the different proposed cost functions.

\paragraph{General case}
\label{sssec:resoptimgal}

We have applied the optimization method on the real FB introduced in Section~\ref{subsec:trnsfglt} with parameters: $N=8$, $k=4$ and $k'=2$. The employed cost function is $\widetilde{J_{\textsf{t}}}$. The result is shown in Figure~\ref{fig:repimpulgltoptim}(b), the coefficients before optimization (obtained with the pseudo-inverse method) are also displayed. It is clear that the impulse responses of optimized filters are better concentrated around the middle of the support. Figure~\ref{fig:repfreqgltoptim} illustrates  that the gain in time localization 
does not entail a too severe 
loss in frequency selectivity of the optimized filters.

A second optimization example is given using an MCLT FB with analysis window $h_{\textsf{a}_{1}}$ and parameters $N=8$, $k=3$ and $k'=7/4$. The resulting  frequency responses after optimization with cost function $\widetilde{J_{\textsf{f}}}$ are represented in Figure~\ref{fig:repfreqBdFoptimsin}(b). 
We observe that the frequency responses after optimization exhibit more regularity and an improved selectivity. 
The proposed cost functions take into account all  synthesis filters at once. It therefore interesting to look more closely at each filter independently and determine whether the optimization leads to better results. In Table \ref{tab:tpsoptimcrit}, the frequency dispersion of each filter is reported before and after optimization with cost function $\widetilde{J_{\textsf{f}}}$. In this case, the overall frequency dispersion of the optimized filters has been noticeably improved and spread variability has been drastically reduced.

\begin{table}[htbp]
\begin{center}
\caption{Frequency dispersion of the synthesis filters optimized with cost function $\widetilde{J_{\mathsf{f}}}$.}
\label{tab:tpsoptimcrit}
\begin{tabular}{|c|c||c|c|}
\hline
Filter & Freq. disp. & Filter & Freq. disp.\\
\hline
\hline
$\widetilde{h}_{0}$ & $0.0290$ & $\widetilde{h}_{0}^{opt}$ & $0.0111$\\
$\widetilde{h}_{1}$ & $0.0851$ & $\widetilde{h}_{1}^{opt}$ & $0.0110$\\
$\widetilde{h}_{2}$ & $0.0569$ & $\widetilde{h}_{2}^{opt}$ & $0.0109$\\
$\widetilde{h}_{3}$ & $0.0606$ & $\widetilde{h}_{3}^{opt}$ & $0.0112$\\
$\widetilde{h}_{4}$ & $0.0658$ & $\widetilde{h}_{4}^{opt}$ & $0.0110$\\
$\widetilde{h}_{5}$ & $0.0596$ & $\widetilde{h}_{5}^{opt}$ & $0.0109$\\
$\widetilde{h}_{6}$ & $0.0363$ & $\widetilde{h}_{6}^{opt}$ & $0.0111$\\
$\widetilde{h}_{7}$ & $0.0156$ & $\widetilde{h}_{7}^{opt}$ & $0.0111$\\
$\widetilde{h}_{8}$ & $0.0411$ & $\widetilde{h}_{8}^{opt}$ & $0.0109$\\
$\widetilde{h}_{9}$ & $0.0231$ & $\widetilde{h}_{9}^{opt}$ & $0.0110$\\
$\widetilde{h}_{10}$ & $0.0508$ & $\widetilde{h}_{10}^{opt}$ & $0.0112$\\
$\widetilde{h}_{11}$ & $0.0499$ & $\widetilde{h}_{11}^{opt}$ & $0.0110$\\
$\widetilde{h}_{12}$ & $0.0574$ & $\widetilde{h}_{12}^{opt}$ & $0.0109$\\
$\widetilde{h}_{13}$ & $0.0521$ & $\widetilde{h}_{13}^{opt}$ & $0.0111$\\
\hline 
Sum   & $0.6833$ &   & $0.1544$\\
\hline
\end{tabular}
\end{center}
\end{table}

\paragraph{Results in the symmetric case}

\begin{figure*}[ht]
\begin{center}
\begin{tabular}{c|c|c}
\includegraphics[height=9cm]{images/kaiser_rep_impul_PIsym.eps}&
\includegraphics[height=9cm]{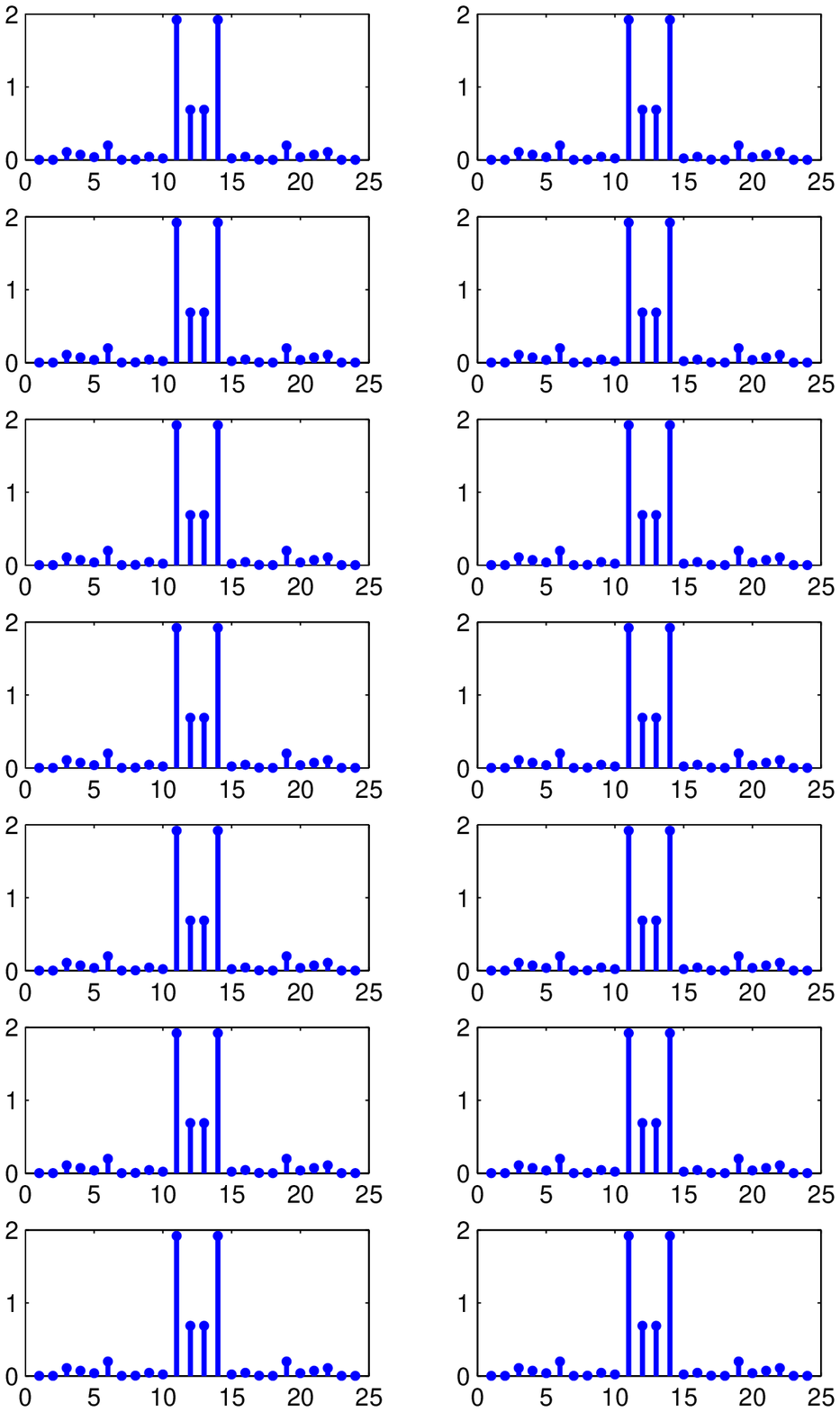}&
\includegraphics[height=9cm]{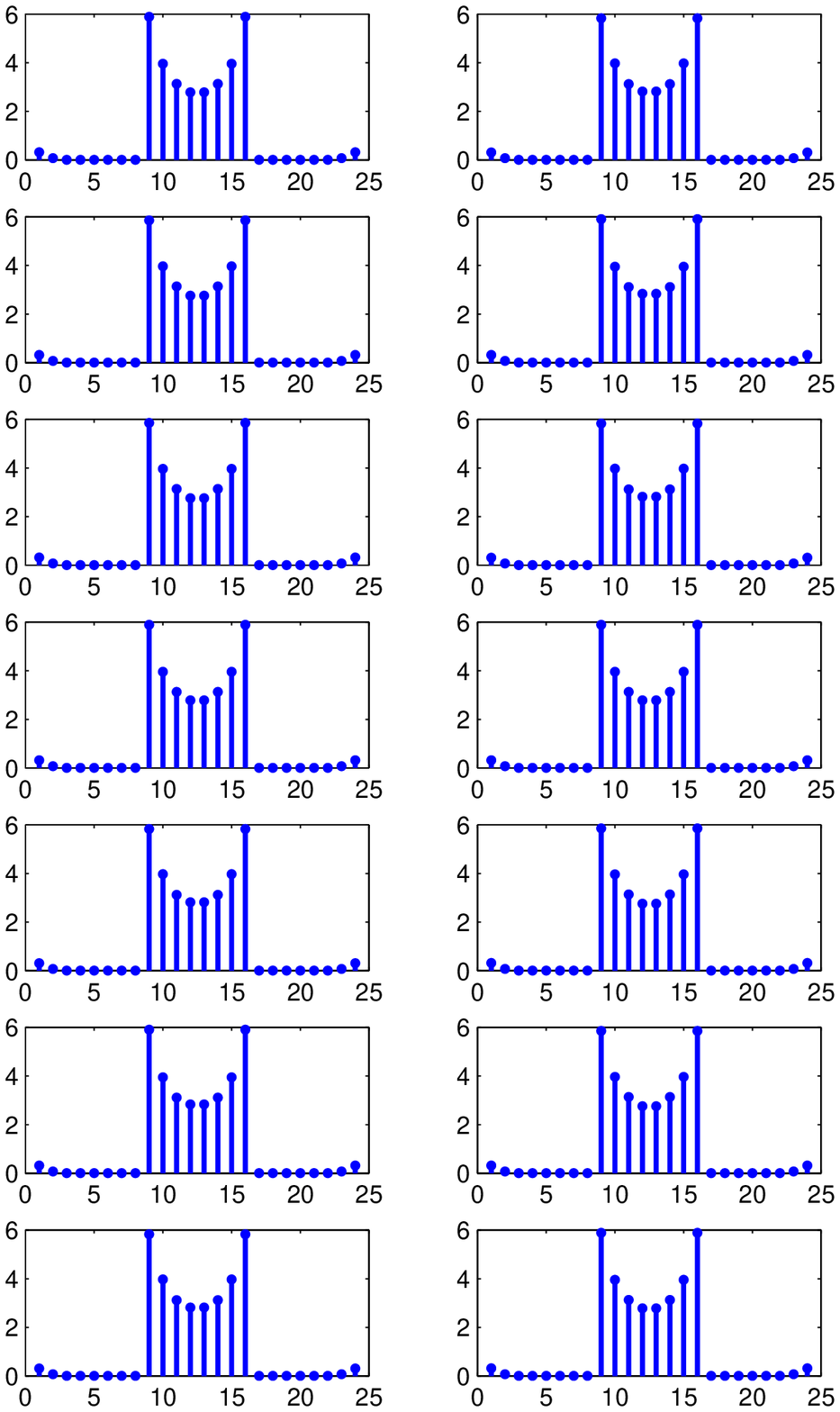}\\
(a)&(b)&(c)\\
\end{tabular}
\caption{Third example (in the MCLT case with window $h_{\textsf{a}_{2}}$): modulus of the impulse responses of synthesis filters (a) before and  after optimization (b) with the cost function $\widetilde{J_{\textsf{ts}}}$ and (c) with $\widetilde{J_{\textsf{fs}}}$.}
\label{fig:repimpulBdFoptimkaiser}
\end{center}
\end{figure*}

\begin{figure*}[ht]
\begin{center}
\begin{tabular}{c|c|c}
\includegraphics[height=9cm]{images/kaiser_rep_freq_PIsym.eps}&
\includegraphics[height=9cm]{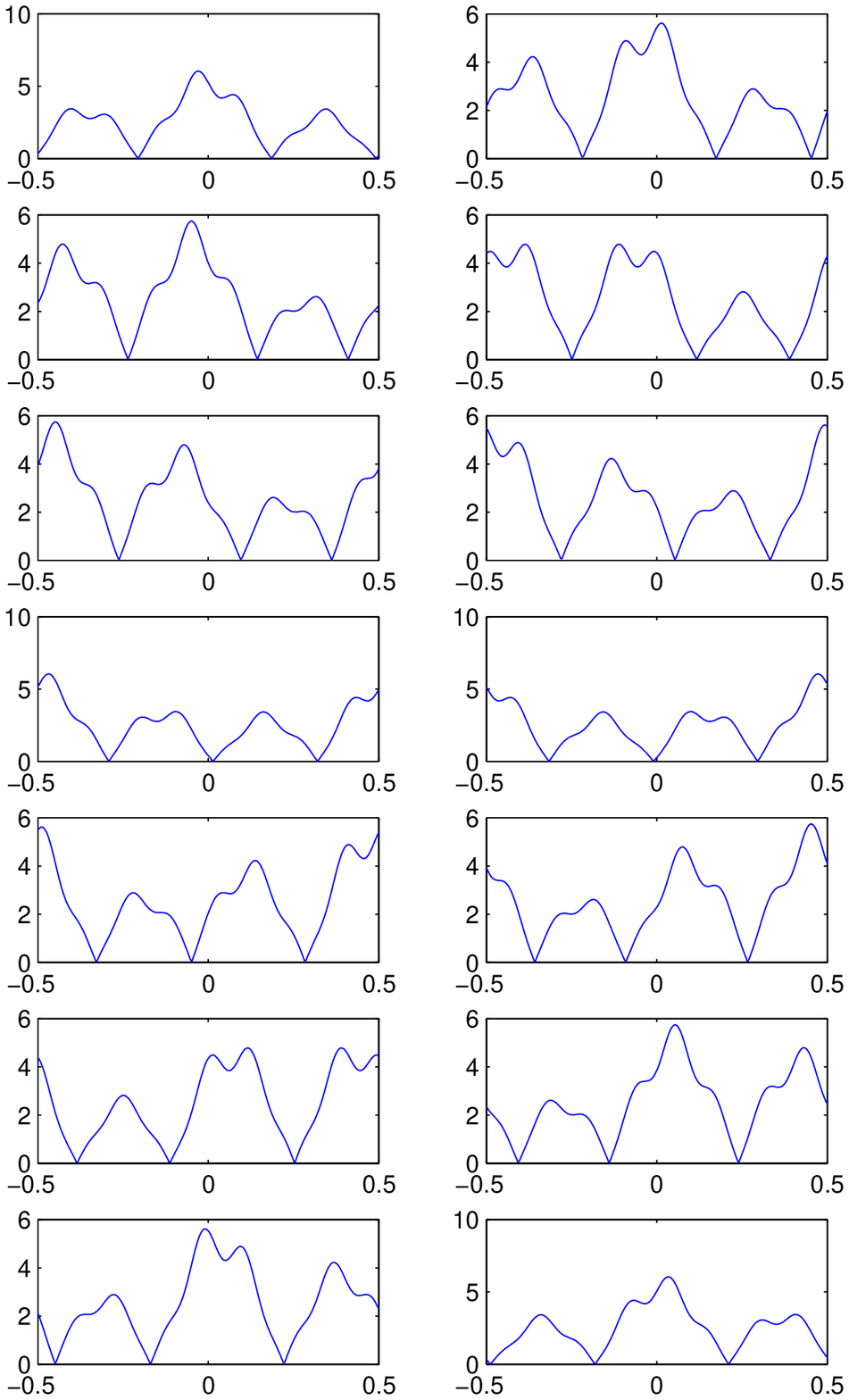}&
\includegraphics[height=9cm]{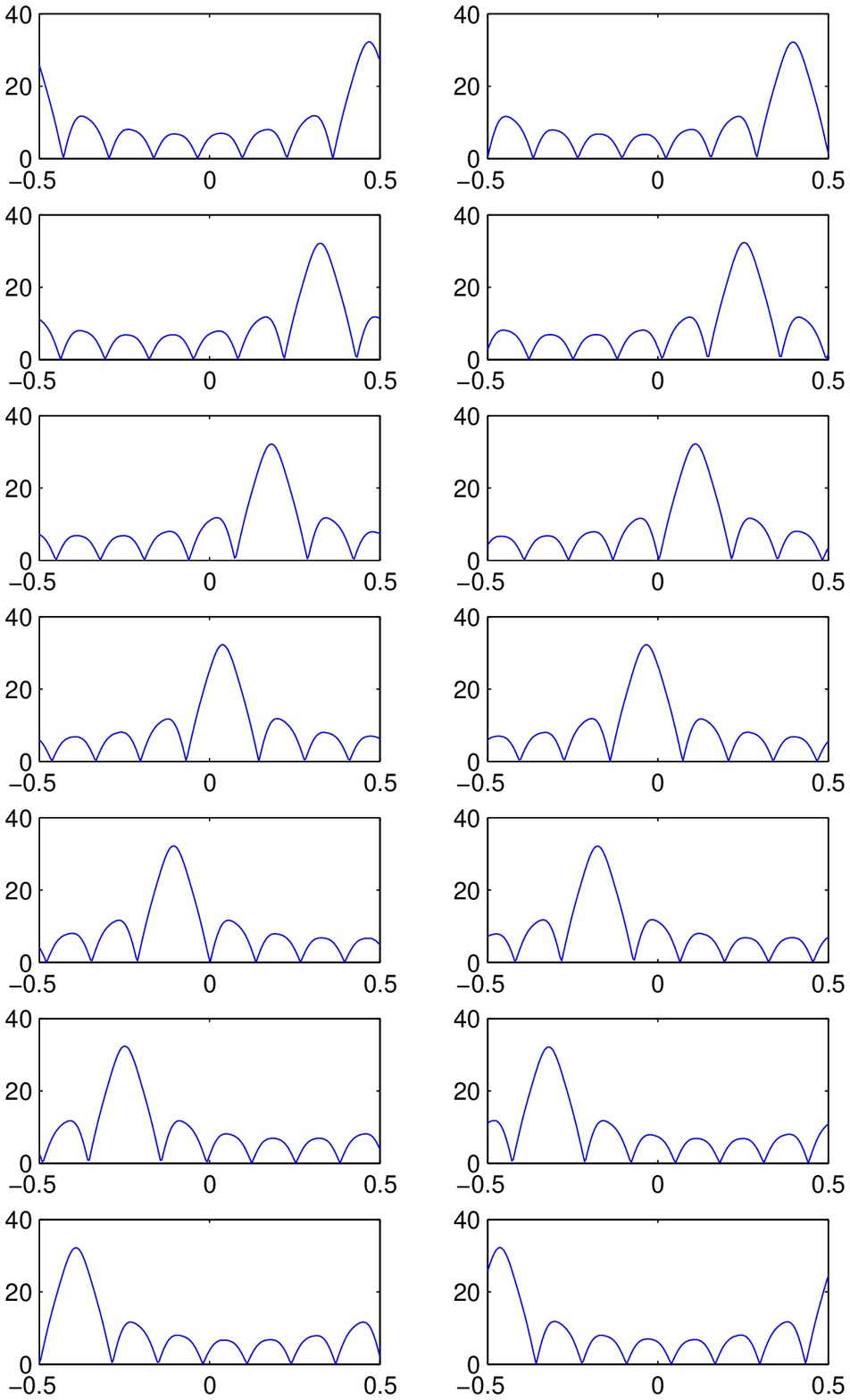}\\
(a)&(b)&(c)\\
\end{tabular}
\caption{Third example (in the MCLT case with window $h_{\textsf{a}_{2}}$): frequency responses of synthesis filters (a) before and  after optimization (b) with the cost function $\widetilde{J_{\textsf{ts}}}$ and (c) with $\widetilde{J_{\textsf{fs}}}$.}
\label{fig:repfreqBdFoptimkaiser}
\end{center}
\end{figure*}

The optimization procedure was next applied in the HS case to the FB of Section~\ref{subsec:trnsfapp} (with analysis window $h_{\textsf{a}_{2}}$). In this case, the cost functions $\widetilde{J_{\textsf{ts}}}$ and $\widetilde{J_{\textsf{fs}}}$ were employed. Once again, the following parameters were used: $N=8$, $k=3$ and $k'=7/4$. Figures \ref{fig:repimpulBdFoptimkaiser} and \ref{fig:repfreqBdFoptimkaiser} show the optimization results. We observe that the optimizations with these two cost functions lead to FBs with different characteristics: as expected, with $\widetilde{J_{\textsf{ts}}}$ the impulse responses are better concentrated than with $\widetilde{J_{\textsf{fs}}}$ and, conversely, with $\widetilde{J_{\textsf{fs}}}$ the frequency selectivity is better than with $\widetilde{J_{\textsf{ts}}}$. 

\subsubsection{Comparison}
\label{sssec:tspcomp}

To conclude this example section, we propose a comparison with an existing filter bank design. We have chosen to compare our design methodology with the filter banks used in \cite{Malvar_H_1999_icassp_mod_cltaap,Young_R_1993_tip_fre_dmeclt}. In these works, the considered FBs correspond to a modulated complex lapped transform with overlap factor $k=2$ and redundancy $k'=2$. For this application, our choice of $N=8$ results in the filters shown in \cite{Young_R_1993_tip_fre_dmeclt}. The synthesis filter bank is then built with a method equivalent to the weighted overlap-add technique. We have applied the methods proposed in this work to compute an optimized synthesis filter bank using the cost function $\widetilde{J_{\textsf{ts}}}$. 
In Table~\ref{tab:comp}, the time dispersion of each synthesis filter computed as explained in \cite{Young_R_1993_tip_fre_dmeclt} $\left(\widetilde{h}_{j}\right)_{j\in\left\{0,...,15\right\}}$ and after optimization $\left(\widetilde{h}_{j}^{opt}\right)_{j\in\left\{0,...,15\right\}}$ are reported as well as the value of the cost function $\widetilde{J_{\mathsf{ts}}}$. The time dispersion was clearly reduced with the proposed method.

\begin{table}[htbp]
\begin{center}
\caption{Time dispersion of the synthesis filters optimized with cost function $\widetilde{J_{\mathsf{ts}}}$.}
\label{tab:comp}
\begin{tabular}{|c|c||c|c|}
\hline
Filter & Time disp. & Filter & Time disp.\\
\hline
\hline
$\widetilde{h}_{0}$, $\widetilde{h}_{15}$ & $21.25$ & $\widetilde{h}_{0}^{opt}$, $\widetilde{h}_{15}^{opt}$ & $2.4152$\\
$\widetilde{h}_{1}$, $\widetilde{h}_{14}$ & $21.25$ & $\widetilde{h}_{1}^{opt}$, $\widetilde{h}_{14}^{opt}$ & $2.1064$\\
$\widetilde{h}_{2}$, $\widetilde{h}_{13}$ & $21.25$ & $\widetilde{h}_{2}^{opt}$, $\widetilde{h}_{13}^{opt}$ & $2.3424$\\
$\widetilde{h}_{3}$, $\widetilde{h}_{12}$ & $21.25$ & $\widetilde{h}_{3}^{opt}$, $\widetilde{h}_{12}^{opt}$ & $2.5516$\\
$\widetilde{h}_{4}$, $\widetilde{h}_{11}$ & $21.25$ & $\widetilde{h}_{4}^{opt}$, $\widetilde{h}_{11}^{opt}$ & $2.5516$\\
$\widetilde{h}_{5}$, $\widetilde{h}_{10}$ & $21.25$ & $\widetilde{h}_{5}^{opt}$, $\widetilde{h}_{10}^{opt}$ & $2.3424$\\
$\widetilde{h}_{6}$, $\widetilde{h}_{9}$ & $21.25$ & $\widetilde{h}_{6}^{opt}$, $\widetilde{h}_{9}^{opt}$ & $2.1064$\\
$\widetilde{h}_{7}$, $\widetilde{h}_{8}$ & $21.25$ & $\widetilde{h}_{7}^{opt}$, $\widetilde{h}_{8}^{opt}$ & $2.4152$\\
\hline 
$\widetilde{J_{\mathsf{ts}}}$  & $340$ &   & $37.6626$\\
\hline
\end{tabular}
\end{center}
\end{table}

\section{Conclusion}

In this paper, we have proposed a method to test that a given oversampled FIR analysis FB is FIR invertible and a method to compute an optimized inverse FB. The optimization was performed for a class of cost functions allowing either to emphasize the time localization or the frequency selectivity of the filters. By rewriting the system defining the synthesis FB, we were able to parameterize the synthesis filters for a given filter length. This parameterization was then used to convert the constrained optimal synthesis problem into an unconstrained one, which can be solved with a simple gradient algorithm. \\
The FB considered here are one-dimensional; it would be interesting to study how the proposed methods could be extended to the multidimensional case. Another perspective could be to study the case of FBs admitting an IIR left inverse that can be approximated using a FIR FB with very long support.

\appendices

\section{Expression of the gradient}
\label{subsec:gradgal}
In this first appendix, we study the gradient of $\widetilde{J}$ (as defined in Section~\ref{sssec:tspgencf}) with respect to $\mathcal{C}$.  We first need to calculate the gradient of $f(\mathcal{C})=\left\| \bi{V}_j\mathcal{C}+\widetilde{\bi{H}}^{0}_{j} \right\|^{2}_{K_j}$. 
The matrix $\mathcal{C}$ being complex, we have
\begin{align*}
\frac{\partial  f}{\partial \mathcal{C}_{m,n}} &= \frac{\partial  f}{\partial \mathcal{C}^{R}_{m,n}}+\imath \frac{\partial  f}{\partial \mathcal{C}^{I}_{m,n}}\\
&= \sum_{i,i',\ell,\ell'}\left((\bi{V}_{j})_{i,m}\delta_{\ell-n}(\overline{\bi{V}_{j}\mathcal{C}+\widetilde{\bi{H}}^{0}_{j}})_{i',\ell'}+\overline{(\bi{V}_{j})}_{i',m}\delta_{\ell'-n}(\bi{V}_{j}\mathcal{C}+\widetilde{\bi{H}}^{0}_{j})_{i,\ell}\right)K_{j}(i,i',\ell,\ell')\\
&+\imath\left(\imath(\bi{V}_{j})_{i,m}\delta_{\ell-n}(\overline{\bi{V}_{j}\mathcal{C}+\widetilde{\bi{H}}^{0}_{j}})_{i',\ell'}-\imath\overline{(\bi{V}_{j})}_{i',m}\delta_{\ell'-n}(\bi{V}_{j}\mathcal{C}+\widetilde{\bi{H}}^{0}_{j})_{i,\ell}\right)K_{j}(i,i',\ell,\ell')\\
&=2\sum_{i,i',\ell}(\overline{\bi{V}_{j}})_{i',m}(\bi{V}_{j}\mathcal{C}+\widetilde{\bi{H}}^{0}_{j})_{i,\ell}K_{j}(i,i',\ell,n).
\end{align*}
From this result, we deduce that
\begin{equation}
\label{eq:gradcplxgen}
\left(\nabla\widetilde{J}(\mathcal{C})\right)_{m,n}=2\sum^{M-1}_{j=0}\frac{\sum_{i,i',\ell}(\overline{\bi{V}_j})_{i',m}(\bi{V}_j\mathcal{C}+\widetilde{\bi{H}}^{0}_{j})_{i,\ell}(\beta_{j}K_{j}(i,i',\ell,n)-\alpha_{j}\Lambda(i,i',\ell,n))}{\beta_{j}^{2}},
\end{equation}
with $\alpha_{j}=\left\| \bi{V}_j\mathcal{C}+\widetilde{\bi{H}}^{0}_{j} \right\|^{2}_{K_j}$ and $\beta_{j}=\left\| \bi{V}_j\mathcal{C}+\widetilde{\bi{H}}^{0}_{j} \right\|^{2}_{\Lambda}$. 

\section{Examples of gradients}
\label{subsec:genex}

A first example is the calculation of the gradient of the cost function $\widetilde{J_{\textsf{t}}}$.
Applying \eqref{eq:gradcplxgen} to  the kernels $(K^{\textsf{t}}_{j})_{0\le j<M}$, we get:
\begin{align*}
\frac{\partial  f}{\partial \mathcal{C}_{m,n}} &= 2\omega_{\textsf{t},j}
\sum_{i}(\overline{\bi{V}_{j}})_{i,m}(\bi{V}_{j}\mathcal{C}+\widetilde{\bi{H}}^{0}_{j})_{i,n}(\boldsymbol{\Gamma}_{j})_{i,n}\\
&= 2\omega_{\textsf{t},j}\left( \bi{V}_{j}^{*}(\boldsymbol{\Gamma}_{j}\odot(\bi{V}_{j}\mathcal{C}+\widetilde{\bi{H}}^{0}_{j}))\right)_{m,n},
\end{align*}
where $(\boldsymbol{\Gamma}_{j})_{i,\ell+p_{1}}=\left|\ell N-i- \overline{m}_j\right|^{\alpha}$ for all $0\le i<N$ and $-p_{1}<\ell\le p_{2}$. The symbol $\odot$ represents the Hadamard product (or pointwise matrix product). Finally we obtain:
$$
\nabla\widetilde{J_{\textsf{t}}}(\mathcal{C})= 2\sum^{M-1}_{j=0}\frac{\bi{V}_j^{*}((\omega_{\textsf{t},j}\beta_{j}\boldsymbol{\Gamma}_{j}-\alpha_{j}\bi{1})\odot(\bi{V}_j\mathcal{C}+\widetilde{\bi{H}}^{0}_{j}))}{\beta_{j}^{2}},
$$
where $\bi{1}_{i,\ell}=1$ for all $0\le i<N$ and $0\le\ell< p$.\\

The same study can be carried out for the cost function $\widetilde{J_{\textsf{f}}}$, with kernels $(K^{\textsf{f}}_{j})_{0\le j<M}$. Rewriting the result under a matrix form does not  simplify the final expression in this case. Hence, the gradient reads:
$$\left(\nabla\widetilde{J_{\textsf{f}}}(\mathcal{C})\right)_{m,n}=2\sum^{M-1}_{j=0}\frac{\sum_{i,i',\ell}(\overline{\bi{V}_j})_{i',m}(\bi{V}_j\mathcal{C}+\widetilde{\bi{H}}^{0}_{j})_{i,\ell}(\beta_{j}K^{\textsf{f}}_{j}(i,i',\ell,n)-\alpha_{j}\delta_{i-i'}\delta_{\ell-n})}{\beta_{j}^{2}}.$$

\section{Gradient functions in the HS case}
\label{ssec:gradfcnsym}

Similarly to Appendix~\ref{subsec:gradgal}, we first compute the gradient of 
$f(\mathcal{C})=\left\| \bi{W}_j\mathcal{C}+\widetilde{\bi{H}}^{0}_{j} \right\|^{2}_{K_j}$ with respect to the matrix $\mathcal{C}$, the only difference being that the matrix $\mathcal{C}$ is now real:
$$
\frac{\partial  f}{\partial \mathcal{C}_{m,n}} 
= \sum_{i,i',\ell,\ell'}\left((\bi{W}_{j})_{i,m}\delta_{\ell-n}(\overline{\bi{W}_{j}\mathcal{C}+\widetilde{\bi{H}}^{0}_{j}})_{i',\ell'}+(\overline{\bi{W}_{j}})_{i',m}\delta_{\ell'-n}(\bi{W}_{j}\mathcal{C}+\widetilde{\bi{H}}^{0}_{j})_{i,\ell}\right)K_{j}(i,i',\ell,\ell').$$
By using this expression and the relation $K_{j}(i,i',\ell,\ell')=\overline{K_{j}(i',i,\ell',\ell)}$, we deduce the gradient of the cost function $\widetilde{J_{\textsf{s}}}$:
\begin{align}
\left(\nabla\widetilde{J_{\textsf{s}}}(\mathcal{C})\right)_{m,n}
&=2\Ree \left(\sum^{M-1}_{j=0}\frac{\sum_{i,i',\ell'}(\bi{W}_j)_{i,m}(\overline{\bi{W}_j\mathcal{C}+\widetilde{\bi{H}}^{0}_{j}})_{i',\ell'}(\beta_{j}K_{j}(i,i',n,\ell')-\alpha_{j}\Lambda(i,i',n,\ell'))}{\beta_{j}^{2}}\right)\label{eq:gradrealgen},
\end{align}
with $\alpha_{j}=\left\| \bi{W}_j\mathcal{C}+\widetilde{\bi{H}}^{0}_{j} \right\|^{2}_{K_j}$ and $\beta_{j}=\left\| \bi{W}_j\mathcal{C}+\widetilde{\bi{H}}^{0}_{j} \right\|^{2}_{\Lambda}$, for all $0\le j<M$. 
Using \eqref{eq:gradrealgen}, the calculation of the gradient of $\widetilde{J_{\textsf{ts}}}$ yields
$$
\nabla\widetilde{J_{\textsf{ts}}}(\mathcal{C})= 2\Ree\left(\sum^{M-1}_{j=0}\frac{\bi{W}_j^{*}((\omega_{\textsf{t},j}\beta_{j}\boldsymbol{\Gamma}_{j}-\alpha_{j}\bi{1})\odot(\bi{W}_j\mathcal{C}+\widetilde{\bi{H}}^{0}_{j}))}{\beta_{j}^{2}}\right),
$$
where $\boldsymbol{\Gamma}_{j}$ is defined as in Appendix~\ref{subsec:genex}. 
For the second cost function we find
\begin{align*}
\left(\nabla\widetilde{J_{\textsf{fs}}}(\mathcal{C})\right)_{m,n}&=2\Ree \left(\sum^{M-1}_{j=0}\frac{\sum_{i,i',\ell'}(\bi{W}_j)_{i,m}(\overline{\bi{W}_j\mathcal{C}+\widetilde{\bi{H}}^{0}_{j}})_{i',\ell'}(\beta_{j}K^{\textsf{f}}_{j}(i,i',n,\ell')-\alpha_{j}\delta_{i-i'}\delta_{n-\ell'})}{\beta_{j}^{2}}\right).
\end{align*}

\end{document}